\documentclass[pdflatex,sn-mathphys-ay]{sn-jnl}

\usepackage{graphicx}
\usepackage{amsmath,amssymb,amsfonts}
\usepackage{amsthm}
\usepackage{booktabs}
\usepackage{bbm}
\usepackage{xurl}
\usepackage{tikz}
\usetikzlibrary{bayesnet, positioning, arrows.meta}
\usepackage{multirow}
\usepackage{algorithm}
\usepackage{algpseudocode}
\usepackage{float}

\theoremstyle{thmstyleone}
\newtheorem{proposition}{Proposition}

\begin{document}

\title{A latent space hidden Markov model for clustering views in longitudinal multilayer networks}

\author*[1]{\fnm{Cecilia} \sur{Manente}}\email{cecilia.manente@uniroma1.it}

\author[1]{\fnm{Marco} \sur{Alfò}}\email{marco.alfo@uniroma1.it}

\author[2]{\fnm{Silvia} \sur{D'Angelo}}\email{dangelos@tcd.ie}

\affil*[1]{\orgdiv{School of Statistical Sciences}, \orgname{Sapienza Università di Roma}, \orgaddress{\street{Piazzale Aldo Moro 5}, \postcode{00185}, \city{Rome}, \country{Italy}}}

\affil[2]{\orgdiv{School of Computer Science and Statistics}, \orgname{Trinity College Dublin}, \orgaddress{\street{College Green}, \postcode{D02 PN40}, \city{Dublin}, \country{Ireland}}}

\abstract{We propose a Bayesian latent space model for clustering longitudinal multilayer networks in which network slices are represented by a finite set of recurrent latent regimes, each associated with a regime-specific latent geometric configuration. Layer-specific hidden Markov chains govern regime membership through a parsimonious additive transition model, allowing heterogeneous temporal dynamics across relational dimensions while sharing a common set of latent configurations.
Edge weights are modelled using a Hurdle Negative Binomial distribution, accommodating both excess zeros and overdispersion in weighted networks. An optional sender--receiver extension further captures persistent node-specific heterogeneity.
A simulation study demonstrates that the proposed model can accurately recover both latent regimes and their associated configurations, with recovery performance depending on the degree of regimes' similarity. We also assess the robustness of the proposed method under model misspecification. Applications to GDELT geopolitical event and European Parliament roll-call voting data demonstrate the flexibility of the framework across diverse domains.}


\maketitle

\section{Introduction}
\label{sec:introduction}
Relational data increasingly take the form of collections of networks on a common set of nodes, observed across time points, relational dimensions, or both; networks at consecutive time points tend to be dependent, as structural patterns persist or recur, while different layers may share part of their structure and differ in overall intensity.
Often a few distinct configurations recur across time points and layers, which motivates clustering the networks into a finite number of latent structural regimes, each a configuration shared by several networks.
Analysing each network separately ignores this shared structure, may produce unstable estimates for sparse networks, and prevents formal inference on regime recurrence.
Latent space models \citep*{hoff2002latent} embed the nodes of a network in a low-dimensional Euclidean space, with edge probabilities or intensities governed by distances between node positions that enter a generalised linear predictor, so weighted and directed networks are handled directly.
\citet*{handcock2007model} extend this representation to model-based clustering of the latent positions, and dynamic extensions let the positions evolve over time \citep{sewell2015latent}; \citet*{kim2018review} review dynamic network models with latent variables.
Multilayer extensions differ in how much latent geometry is common across layers: \citet{gollini2016joint} generate all layers from one configuration with layer-specific intercepts, \citet*{dangelo2019latent} let each layer rescale the shared distances, changing how strongly geometry drives connection, and \citet{salter2017multiview} allow layer-specific positions and model the dependence between layers directly.
Other proposals estimate from the data how much structure is shared: \citet*{macdonald2022latent} combine common and layer-specific components, and \citet{Clarke2025} develop a latent position co-clustering model for multiplex networks.
None of these models temporal dependence or recurrent regimes.
Temporal and multilayer structure have also been modelled jointly: \citet{Loyal2023} propose a Bayesian nonparametric latent space model for evolving community structure in dynamic networks, \citet{Loyal20232} an eigenmodel for dynamic multilayer networks with smoothly time-varying latent positions, \citet*{baum2024doubly} a dynamic multiplex random dot product graph estimated by spectral embedding, and \citet*{lan2025tensor} a tensor state space model based on a Tucker decomposition of a latent edge-probability tensor.
These methods describe evolution through time-varying latent representations or low-rank decompositions, not through a finite set of configurations that persist and reappear, and most assume binary edges.
Closer to our proposal, \citet{park2020detecting} detect structural changes through a hidden Markov extension of multilinear tensor regression \citep{hoff2015multilinear}, but their regimes follow a change-point structure and cannot reappear once left.
The population-of-networks framework \citep*{durante2017nonparametric, signorelli2020model} represents a collection of networks on a common node set through a few latent structural templates, which suits settings with many networks relative to the genuinely distinct configurations, but treats the collection as exchangeable and so ignores temporal dependence.
Outside latent space models, dynamic stochastic block models \citep*{matias2017statistical, ludkin2018dynamic} let a latent Markov chain drive the group memberships of the nodes, modelling changes in community membership rather than recurrent global configurations.
Temporal exponential random graph models \citep*{hanneke2010discrete, leifeld2018temporal} extend exponential random graph models to sequences of networks through temporally lagged terms, so dependence enters through network statistics rather than latent structural states.
Tensor decompositions \citep{kolda2009tensor} and clustering of network summary statistics \citep*{onnela2012taxonomies} reveal common patterns across networks, but are essentially descriptive and ignore temporal dependence.
None of these approaches combines, within a single generative model for weighted networks, (i) a finite set of latent configurations that can recur over time, (ii) Markov dynamics on the regime labels, and (iii) layer-specific transition dynamics over a shared set of configurations.
We propose a Bayesian latent space model for longitudinal multilayer networks that represents the collection through a finite number of recurrent latent regimes, each with a time-invariant latent embedding of the common node set shared by the networks assigned to it.
Layer-specific hidden Markov chains \citep*{rabiner1989tutorial, bartolucci2013latent} govern regime transitions, so layers may follow different temporal dynamics while drawing on a common set of configurations; given the regime sequences, the networks are independent.
An additive parameterisation of the transition matrices separates a component shared by all layers from layer-specific deviations, keeping the number of parameters under control, and edge weights follow a Hurdle Negative Binomial distribution, which accommodates the excess zeros and overdispersion typical of count-valued networks, and an optional sender--receiver extension captures persistent node-specific heterogeneity.
Networks in the same regime share one configuration, so information is pooled across time points and layers, and the posterior distribution of the regime sequences yields a dynamic clustering of the networks, suited to settings where network structure switches among a few recurrent configurations rather than evolving smoothly.

Section~\ref{sec:model} presents the model and Section~\ref{sec:inference} posterior inference; Sections~\ref{sec:simulation} and~\ref{sec:applications} report the simulation study and the applications to GDELT event counts and European Parliament roll-call votes, and Section~\ref{sec:discussion} discusses limitations and extensions.

\section{Model}
\label{sec:model}

\subsection{Data and notation}
\label{sec:notation}
We observe networks on a common set of $n$ nodes at $T$ time points describing $K$ relational layers, with complete collection $\mathcal{Y}=\{\mathbf{Y}_{tk}:t=1,\ldots,T,\;k=1,\ldots,K\}$.
Each $\mathbf{Y}_{tk}$ is an $n\times n$ matrix whose $(i,j)$-th entry, $Y_{tk,ij}$, is the value of the edge from node $i$ to node $j$ at time $t$ in layer $k$, with $Y_{tk,ii}=0$ for all $i$; entries may be binary or non-negative integer counts, as in flow or interaction data.
The pair $(t,k)$ is a network slice, and there are $KT$ slices.
We focus on directed networks; for undirected ones, sums over ordered pairs $i\neq j$ become sums over $i<j$.

\subsection{Latent space representation}
\label{sec:lsm}
Following \citet{hoff2002latent}, nodes are embedded in a low-dimensional Euclidean space, where the propensity to interact decreases with latent distance.
The $KT$ networks arise from $G$ latent configurations, with $G$ substantially smaller than $KT$, each possibly shared by slices at different times, in different layers or both.
For each regime $g=1,\ldots,G$, let $\mathbf{Z}_g\in\mathbb{R}^{n\times d}$ denote the matrix of latent node coordinates, where $\boldsymbol{z}_{ig}=(z_{ig,1},\ldots,z_{ig,d})^\top\in\mathbb{R}^d$ is the position of node $i$ under regime $g$.
The Euclidean distance between nodes $i$ and $j$ within regime $g$ is $D_g(i,j) = 
\|\boldsymbol{z}_{ig}-\boldsymbol{z}_{jg}\|_2 =
\sqrt{\textstyle\sum_{\ell=1}^d (z_{ig,\ell}-z_{jg,\ell})^2}.
$
The configurations are regime-specific but time-invariant: temporal variation enters only through the regime process of Section~\ref{sec:hmm}, so a configuration can reappear at later times or in other layers.
The positions have independent Gaussian priors,
$\boldsymbol{z}_{ig}\overset{\text{iid}}{\sim}\mathcal{N}_d(\mathbf{0},\sigma_z^2\mathbf{I}_d)$ for
$i=1,\ldots,n$ and $g=1,\ldots,G$, with fixed $\sigma_z^2$.

\paragraph{Translation and rotation invariance.}
The likelihood depends on $\mathbf{Z}_g$ only through pairwise distances, so it is invariant to translations, rotations and reflections.
Write $\mathbf{Z}_g=\mathbf{W}_g+\mathbf{1}_n\bar{\boldsymbol{z}}_g^\top$, with centroid $\bar{\boldsymbol{z}}_g=n^{-1}\sum_{i} \boldsymbol{z}_{ig}$ and centred configuration $\mathbf{W}_g$.
Under the iid Gaussian prior, $\mathbf{W}_g$ and $\bar{\boldsymbol{z}}_g$ are independent, with $\bar{\boldsymbol{z}}_g\sim\mathcal{N}_d(\mathbf{0},\sigma_z^2n^{-1}\mathbf{I}_d)$. Hence, the likelihood depends only on $\mathbf{W}_g$, $p(\mathbf{W}_g,\bar{\boldsymbol{z}}_g\mid\mathcal{Y})=p(\mathbf{W}_g\mid\mathcal{Y})\,p(\bar{\boldsymbol{z}}_g)$.
The data thus carry no information on the centroid: the sampler draws it from its prior (Section~\ref{sec:mcmc}), and configurations are summarised by recentring the saved draws.
Rotations and reflections, that are prior-invariant, are not identified and are removed by orthogonal Procrustes alignment \citep[Ch.~7]{dryden2016statistical} (Section~\ref{sec:relabeling}).

\subsection{Emission model}
\label{sec:emission}
Given the latent positions and $S_{tk}=g$, the edge values $\{Y_{tk,ij}\}_{i\neq j}$ are conditionally independent with emission distribution $f(\cdot)$,
\begin{equation}
Y_{tk,ij}\mid S_{tk}=g,\mathbf{Z}_g,\alpha_{tk},\boldsymbol{\psi}_{tk}
\;\sim\;
f\!\left(\,\cdot \mid \alpha_{tk}-D_g(i,j),\;\boldsymbol{\psi}_{tk}\right),
\label{eq:link}
\end{equation}
where the network-specific intercept $\alpha_{tk}\in\mathbb{R}$ sets the overall connectivity of slice $(t,k)$, and $\boldsymbol{\psi}_{tk}$ collects the remaining distribution-specific parameters, independently of the dyad and the regime.
The linear predictor $\alpha_{tk}-D_g(i,j)$ enters $f(\cdot)$ only through its intensity component. Therefore, slices in the same regime share their geometry and differ only in connectivity, intensity and dispersion.
The log-likelihood of slice $(t,k)$ under regime $g$ is $
\ell(\mathbf{Y}_{tk}\mid g,\alpha_{tk},\boldsymbol{\psi}_{tk})
=
\sum_{i\neq j}
\log f\!\left(Y_{tk,ij} \mid \alpha_{tk}-D_g(i,j),\;\boldsymbol{\psi}_{tk}\right).
$
This formulation accommodates Bernoulli emissions for binary networks, Poisson or Negative Binomial for counts and log-Normal for positive continuous weights. Here, we adopt the Hurdle Negative Binomial model, suited for directed count networks with substantial overdispersion and many zeros.

\paragraph{Identifiability of the network-specific intercepts.}
Beyond the invariances of Section~\ref{sec:lsm}, there is an indeterminacy between the intercepts and the scale of the latent distances, since Equation~\eqref{eq:link} depends on $\alpha_{tk}$ and $\mathbf{Z}_g$ only through $\alpha_{tk}-D_g(i,j)$.
They are not strictly confounded, as adding the same constant to every pairwise distance does not in general give a configuration realisable in $\mathbb{R}^d$, but a shift in $\alpha_{tk}$ can be approximately compensated by the distances, so $\alpha_{tk}$ and the mean distance $\bar D_{S_{tk}}$ are strongly correlated in the posterior.
The proper Gaussian prior on the positions regularises the scale of $D_g$, but $\alpha_{tk}$ remains coupled to it, so we report the centred-intercept contrast
\begin{equation}
\tilde\alpha_{tk}
=
\alpha_{tk}-\bar D_{S_{tk}},
\qquad
\bar D_g=\frac{1}{n(n-1)}\sum_{i\neq j} D_g(i,j),
\label{eq:alpha_contrast}
\end{equation}
rather than $\alpha_{tk}$ alone.
Equivalently, $\alpha_{tk}-D_g(i,j) = \tilde\alpha_{tk}-\{D_g(i,j)-\bar D_g\}$, which separates the overall network level from the variation of the distances around their mean.

\subsubsection{The Hurdle Negative Binomial emission}
\label{sec:emission_hnb}
The Hurdle Negative Binomial (HNB) distribution combines a Bernoulli component for whether an edge is active with a zero-truncated Negative Binomial for the magnitude of positive values, separating presence from magnitude \citep{mullahy1986specification}. We follow the specification of \citet*{manente2026dynamiclatentspacemodel} for dynamic single-layer networks.
Let $
\mathrm{NB}(y\mid\mu,r)
=
\frac{\Gamma(y+r)}{\Gamma(r)\,\Gamma(y+1)}
\left(\frac{r}{r+\mu}\right)^{r}
\left(\frac{\mu}{r+\mu}\right)^{y}
$
be the Negative Binomial mass function with mean $\mu>0$ and dispersion $r>0$; smaller values of $r$ mean stronger overdispersion, and $\mathrm{NB}(y\mid\mu,r)$ converges to the Poisson mass function with mean $\mu$ as $r\to\infty$.
The edge values are distributed as
\begin{equation}
P(Y_{tk,ij}=y\mid\pi_{tk},\mu_{tk,ij},r_{tk})
=
\begin{cases}
1-\pi_{tk}, & y=0, \\[6pt]
\displaystyle
\pi_{tk}\,
\frac{\mathrm{NB}(y\mid\mu_{tk,ij},r_{tk})}
{1-\mathrm{NB}(0\mid\mu_{tk,ij},r_{tk})},
& y>0,
\end{cases}
\label{eq:hnb}
\end{equation}
where $\pi_{tk}\in(0,1)$ is the probability that an edge is active, and $\mu_{tk,ij}$ and $r_{tk}$ are the mean and dispersion of the untruncated component, with
\begin{equation}
\mu_{tk,ij}
=
\exp\!\left\{\alpha_{tk}-D_{S_{tk}}(i,j)\right\},
\label{eq:mu}
\end{equation}
so that nodes closer in the latent space generate larger expected flows.
The conditional mean of the positive counts,
$E[Y_{tk,ij}\mid Y_{tk,ij}>0] = \mu_{tk,ij}\,\{1-\mathrm{NB}(0\mid\mu_{tk,ij},r_{tk})\}^{-1}$, exceeds $\mu_{tk,ij}$.
The hurdle probability is $\pi_{tk}=\{1+\exp(-\beta_{tk})\}^{-1}$, with $\beta_{tk}\in\mathbb{R}$, so that $\boldsymbol{\psi}_{tk}=(\beta_{tk},r_{tk})$; as in \citet{manente2026dynamiclatentspacemodel}, $\pi_{tk}$ is common to all dyads within a slice, so the latent geometry governs the magnitude of interactions, not their occurrence.
The network-specific parameters have independent priors
$\alpha_{tk}\overset{\text{iid}}{\sim}\mathcal{N}(\mu_\alpha,\sigma_\alpha^2)$
and
$\beta_{tk}\overset{\text{iid}}{\sim}\mathcal{N}(\mu_\beta,\sigma_\beta^2)$; for the dispersion we set $a_{tk}=r_{tk}^{-1/2}$, with half-Normal prior
$a_{tk}\overset{\text{iid}}{\sim}\mathcal{N}^{+}(0,\sigma_a^2)$ \citep{gelman2006prior}, which is proper and has its mode at $a_{tk}=0$, the Poisson limit, so that overdispersion is introduced only when strongly supported by the data.

\subsection{Hidden Markov structure and transition model}
\label{sec:hmm}
A hidden state $S_{tk}\in\{1,\ldots,G\}$, specific to slice $(t,k)$, describes the evolution of regime assignments: for each layer $k$, $\{S_{tk}\}_{t=1}^T$ is described by a first-order Markov chain \citep{rabiner1989tutorial,bartolucci2013latent} with transition probabilities, for $t\geq 2$,
\begin{equation}\hspace*{-.2cm}
P(S_{tk}=g\mid S_{t-1,k}=h,S_{t-2,k},\dots)
=
P(S_{tk}=g\mid S_{t-1,k}=h)
=: p_{hg}^{(k)}(t).
\label{eq:markov}
\end{equation}
Assuming time homogeneity, $p_{hg}^{(k)}(t)=p_{hg}^{(k)}$, bounds the number of transition parameters as $T$ grows (see Section~\ref{sec:discussion} for extensions).
A priori, the $K$ chains are independent given the transition parameters, so layers are linked only through the shared baseline $\boldsymbol{\Gamma}$ and configurations $\{\mathbf{Z}_g\}_{g=1}^G$, and regime changes need not be synchronised.
The initial state distribution is uniform, $\omega_g := P(S_{1k}=g)=G^{-1}$ for all $k$ and $g$.

\paragraph{Additive parameterisation of the transition probabilities.}
For each layer $k$, a row-wise softmax of the utilities $\eta_{hg}^{(k)}$ gives the transition probabilities $p_{hg}^{(k)}=\exp\{\eta_{hg}^{(k)}\}/\sum_{g'=1}^{G}\exp\{\eta_{hg'}^{(k)}\}$, $h,g=1,\ldots,G$, which form the transition matrix $\mathbf{P}^{(k)}$ of layer $k$.
Instead of a separate matrix for every layer, each utility is the sum of a shared baseline and a layer-specific deviation, $\eta_{hg}^{(k)}=\gamma_{hg}+\lambda_{kg}$, where $\gamma_{hg}$ is the baseline utility of moving from regime $h$ to regime $g$ and $\lambda_{kg}$ the propensity of layer $k$ to move towards regime $g$, whatever the origin.
The matrix $\boldsymbol{\Gamma}=(\gamma_{hg})$ captures the dynamics common to all layers, with persistence on its diagonal and baseline switching off it. The matrix $\boldsymbol{\Lambda}=(\lambda_{kg})$ makes each destination more or less attractive for individual layers.

\paragraph{Identifiability of the transition parameters.}
We take regime $G$ as the reference destination, $\gamma_{hG}=\lambda_{kG}=0$, with independent priors $\gamma_{hg}\sim\mathcal{N}(0,\sigma_\gamma^2)$ and
$\lambda_{kg}\sim\mathcal{N}(0,\sigma_\lambda^2)$ for $g<G$.
Since $\eta_{hg}^{(k)}=\gamma_{hg}+\lambda_{kg}$ is unchanged by $\gamma_{hg}\mapsto\gamma_{hg}+c_g$, $\lambda_{kg}\mapsto\lambda_{kg}-c_g$, only $\eta_{hg}^{(k)}$ and $p_{hg}^{(k)}$ are identified: $(G-1)(G+K-1)$ parameters, against $KG(G-1)$ for a fully layer-specific parameterisation, a substantial reduction for large $K$.
On each saved draw we compute the canonical representative with layer deviations summing to zero,
\begin{equation}
\gamma_{hg}^{\mathrm{c}} = \frac{1}{K}\sum_{k=1}^K\eta_{hg}^{(k)},
\qquad
\lambda_{kg}^{\mathrm{c}} = \lambda_{kg}-\frac{1}{K}\sum_{k'=1}^K\lambda_{k'g},
\qquad g<G,
\label{eq:gamma_delta_identified}
\end{equation}
that is, the average utility across layers and the deviation for layer $k$; the non-identified direction is handled within the sampler (Section~\ref{sec:mcmc}).

\paragraph{Label switching.}
The likelihood is invariant to permutations of the regime labels $1,\ldots,G$, so the posterior has up to $G!$ equivalent modes \citep{stephens2000dealing}, and regime-specific quantities such as $\mathbf{Z}_g$ or $\gamma_{hg}$ can be averaged only after aligning the draws to a common labelling \citep*{celeux2000computational} (Section~\ref{sec:relabeling}); $G$ is treated as fixed and selected as described in Section~\ref{sec:model_selection}.

\subsection{Regime separation}
\label{sec:separation}
We measure the separation between regimes $g$ and $h$ by the expected log-likelihood gap for a slice $\mathbf{Y}_{tk}$ generated under regime $g$, with $\alpha_{tk}$ and $\boldsymbol{\psi}_{tk}$ held fixed,
\begin{equation}
\begin{aligned}
\Delta_{gh}
&=
\mathbb{E}_{g}\!\left[
\ell(\mathbf{Y}_{tk}\mid g,\alpha_{tk},\boldsymbol{\psi}_{tk})
-\ell(\mathbf{Y}_{tk}\mid h,\alpha_{tk},\boldsymbol{\psi}_{tk})
\right] \\
&=
D_{\mathrm{KL}}\!\left[
p(\cdot\mid \mathbf{Z}_g,\alpha_{tk},\boldsymbol{\psi}_{tk})
\,\middle\|\,
p(\cdot\mid \mathbf{Z}_h,\alpha_{tk},\boldsymbol{\psi}_{tk})
\right],
\end{aligned}
\label{eq:delta}
\end{equation}
that is, the Kullback--Leibler divergence between the emission distributions of the two regimes at the same network-specific parameters.
Since $D_{\mathrm{KL}}\geq 0$, with equality if and only if the two distributions coincide almost everywhere \citep[Theorem~2.6.3]{cover1999elements}, large values of $\Delta_{gh}$ indicate substantially different distributions over networks, and values close to zero regimes that are difficult to distinguish from the data.
Under the HNB emission, the divergence has a closed form that isolates the contribution of each component.

\begin{proposition}[Separation under the HNB emission]
\label{cor:delta_pi}
Under Equations~\eqref{eq:hnb}--\eqref{eq:mu}, and writing
$\mathrm{ZTNB}(\mu,r)$ for the Negative Binomial distribution with mean $\mu$ and dispersion $r$ truncated at zero,
\begin{equation}
\Delta_{gh}
=
\pi_{tk}\sum_{i\neq j}
D_{\mathrm{KL}}\!\left[
\mathrm{ZTNB}(\mu^{g}_{tk,ij},r_{tk})
\,\middle\|\,
\mathrm{ZTNB}(\mu^{h}_{tk,ij},r_{tk})
\right],
\label{eq:delta_pi}
\end{equation}
where $\mu^{g}_{tk,ij}=\exp\{\alpha_{tk}-D_g(i,j)\}$.
\end{proposition}

\begin{proof}
Under both regimes $P(Y_{tk,ij}=0)=1-\pi_{tk}$, since $\pi_{tk}$ depends neither on $(i,j)$ nor on the regime.
The contribution of the zero outcome to the divergence is therefore zero, and the positive outcomes contribute $\pi_{tk}$ times the divergence between the corresponding zero-truncated Negative Binomial distributions.
Summing over the conditionally independent dyads gives Equation~\eqref{eq:delta_pi}.
\end{proof}

Since $\pi_{tk}$ is common to all dyads and regimes, the hurdle component does not contribute to separation: all the information distinguishing regimes lies in the positive edges.
For fixed $\alpha_{tk}$, $r_{tk}$ and configurations, separation is proportional to $\pi_{tk}$, so halving the density of the networks halves the separation between any two regimes, as verified in Section~\ref{sec:simulation}.

\paragraph{Empirical KL-based regime-separation measure.}
In practice $\Delta_{gh}$ is replaced by
\begin{equation}
\widehat{\Delta}_{gh}
=
\frac{1}{|\mathcal{T}_g|}
\sum_{(t,k)\in\mathcal{T}_g}
\left[
\ell(\mathbf{Y}_{tk}\mid \widehat{\mathbf{Z}}_g,\widehat\alpha_{tk},\widehat{\boldsymbol{\psi}}_{tk})
-\ell(\mathbf{Y}_{tk}\mid\widehat{\mathbf{Z}}_h,\widehat\alpha_{tk},\widehat{\boldsymbol{\psi}}_{tk})
\right],
\label{eq:delta_hat}
\end{equation}
where $\mathcal{T}_g=\{(t,k):\widehat S_{tk}=g\}$ is the set of slices with modal regime $g$ and hatted quantities are posterior means, both computed after the post-processing of Section~\ref{sec:relabeling}.
Unlike $\Delta_{gh}$, an expectation under the true distribution of regime $g$, $\widehat{\Delta}_{gh}$ averages over the slices classified to $g$, so it is not a Kullback--Leibler divergence and may be negative, hence the name \emph{empirical KL-based regime-separation measure}.
As the same data assign the slices and evaluate the gap, it is a diagnostic, not an estimator of $\Delta_{gh}$; values are normalised by the number of directed dyads $n(n-1)$ for comparison across network sizes. The quantity
\begin{equation}
\widehat\Delta_{\min}(G)=\min_{g\neq h}\widehat{\Delta}_{gh}
\label{eq:delta_min_def}
\end{equation}
is a diagnostic of regime redundancy: it approaches zero when two fitted regimes are nearly indistinguishable and both occupied, suggesting more regimes than the data support.
It requires all regimes to be occupied: an empty regime has a configuration driven by the prior alone, so the log-likelihood differences against it are large and positive, and $\widehat\Delta_{\min}(G)$ is inflated, precisely when it should detect redundancy.
It is therefore always reported and read together with the minimum regime occupancy (Section~\ref{sec:model_selection}).

\subsection{Sender--receiver extension}
\label{sec:sr}
In Equation~\eqref{eq:mu}, persistent differences in the tendency of nodes to send or receive interactions must be absorbed by the latent geometry.
Following the multiplicative specification by \citet*{d2020modeling}, extended by \citet{manente2026dynamiclatentspacemodel}, we introduce global sender and receiver effects $u_i$ and $v_j$, $i,j=1,\ldots,n$, in the mean of the untruncated Negative Binomial component,
\begin{equation}
\mu_{tk,ij}
=
\exp\!\left[
\alpha_{tk}\frac{u_i+v_j}{2} - D_{S_{tk}}(i,j)
\right].
\label{eq:mu_sr}
\end{equation}
The hurdle probability $\pi_{tk}$ is unchanged, as node heterogeneity is assumed to affect the intensity of interactions rather than the probability that an edge forms.
The effects are constant across time, layers and regimes, so they describe persistent node characteristics, separate from the regime-specific structure in $D_{S_{tk}}$ and unaffected by label switching; multiplication by $\alpha_{tk}$ scales them with the connectivity of each slice.
Their interpretation as propensities requires $\alpha_{tk}>0$, so here $\alpha_{tk}\sim\mathcal{N}^{+}(\mu_\alpha,\sigma_\alpha^2)$, truncated to $(0,\infty)$, replaces the prior of Section~\ref{sec:emission_hnb}.
Equation~\eqref{eq:mu_sr} is invariant to the shift $u_i\leftarrow u_i+c_1$ and $v_j\leftarrow v_j-c_1$ and to the rescaling $\alpha_{tk}^{\ast}=\alpha_{tk}/c_2$, $u_i^{\ast}=c_2\,u_i$, $v_j^{\ast}=c_2\,v_j$, with $c_2\neq0$, including a sign change for $c_2<0$; as in \citet{d2020modeling}, who note these invariances, we fix $u_{i_u}=1$ and $v_{j_v}=1$.
As in \citet{manente2026dynamiclatentspacemodel}, the reference nodes have the median observed out-degree ($i_u$) and in-degree ($j_v$), where the degree of a node counts the observed positive edges $(i,j,t,k)$ in which it is the sender (respectively receiver), over all slices.
The remaining effects have independent priors $u_i\sim\mathcal{N}(0,\sigma_u^2)$, $i\neq i_u$, and $v_j\sim\mathcal{N}(0,\sigma_v^2)$, $j\neq j_v$, with $\sigma_u=\sigma_v=1$.
Since $\alpha_{tk}$ is no longer an additive intercept, the contrast of Equation~\eqref{eq:alpha_contrast} becomes $\tilde\alpha_{tk}=\alpha_{tk}\bar w-\bar D_{S_{tk}}$, with $\bar w=\frac{1}{n(n-1)}\sum_{i\neq j}(u_i+v_j)/2$.

\section{Posterior inference}
\label{sec:inference}

\subsection{MCMC scheme}
\label{sec:mcmc}
A summary of the model is presented in the Appendix  (S1). Posterior inference uses a Metropolis-within-Gibbs algorithm (Algorithm~S1 in the Appendix): the regime sequences of each layer are drawn jointly from their full conditional by forward filtering backward sampling (FFBS; Section~\ref{sec:ffbs}), and the other parameters, whose full conditionals are not available in closed form, by Metropolis--Hastings (MH) random-walk steps.
All chains start from the same deterministic initial values and differ only in their random seeds: the regime sequences are obtained by clustering the slices on their centred log-counts, the latent configurations by classical multidimensional scaling, $\alpha_{tk}$ and $\beta_{tk}$ by moment estimates, with $r_{tk}=1$, $\boldsymbol{\Gamma}=\boldsymbol{\Lambda}=\mathbf{0}$ and, under the sender--receiver extension, $u_i=v_j=1$.

\paragraph{Update of the latent positions $\mathbf{Z}_g$.}
For each regime $g$ and node $i$, the proposal $\boldsymbol{z}_{ig}^{\ast}=\boldsymbol{z}_{ig}+\boldsymbol{\epsilon}$, with $\boldsymbol{\epsilon}\sim\mathcal{N}_d(\mathbf{0},\tau_z^2 \mathbf{I}_d)$, is accepted with probability $\min(1,R_z)$, where
\begin{equation}
R_z=\exp\Bigl\{
\sum_{(t,k):S_{tk}=g}
[\ell(\mathbf{Y}_{tk}\mid \boldsymbol{z}_{ig}^{\ast})-\ell(\mathbf{Y}_{tk}\mid \boldsymbol{z}_{ig})]
+\log\frac{p(\boldsymbol{z}_{ig}^{\ast})}{p(\boldsymbol{z}_{ig})}
\Bigr\},
\label{eq:mh_z}
\end{equation}
and the proposal ratio cancels by symmetry.
An empty regime is updated from the prior alone and remains available to later FFBS updates, as no occupancy constraint is imposed.
Before the node-wise updates, the translation of $\mathbf{Z}_g$ is drawn from its full conditional, $\mathbf{Z}_g\leftarrow \mathbf{Z}_g+\mathbf{1}_n \boldsymbol{c}^\top$ with $\boldsymbol{c}\sim\mathcal{N}_d(-\bar{\boldsymbol{z}}_g,\sigma_z^2 n^{-1} \mathbf{I}_d)$, so the updated centroid $\bar{\boldsymbol{z}}_g+\boldsymbol{c}$ is a draw from $\mathcal{N}_d(\mathbf{0},\sigma_z^2n^{-1}\mathbf{I}_d)$; as the likelihood is constant along translations, this step involves only the prior and leaves the posterior invariant.

\paragraph{Update of the network-specific parameters.}
The parameters $\alpha_{tk}$, $\beta_{tk}$ and $a_{tk}=r_{tk}^{-1/2}$, conditionally independent across slices, are updated in turn by MH random-walk steps analogous to Equation~\eqref{eq:mh_z}, under the current regime $g=S_{tk}$ and the priors of Section~\ref{sec:emission_hnb}.
The parameter $a_{tk}$ and, under the sender--receiver extension, $\alpha_{tk}$ use the reflected proposal $x^\ast=|x+\epsilon|$ on $(0,\infty)$, which is symmetric, so the proposal ratio again cancels.

\paragraph{Update of the sender and receiver effects.}
Under the sender--receiver extension (Section~\ref{sec:sr}), each free sender effect $u_i$, $i\neq i_u$, is updated by proposing $u_i^{\ast}=u_i+\epsilon$, with $\epsilon\sim\mathcal{N}(0,\tau_u^2)$, and accepting with probability $\min(1,R_u)$, where $R_u=\exp\{\sum_{t=1}^T\sum_{k=1}^K\sum_{j\neq i}[\ell(Y_{tk,ij}\mid u_i^{\ast})-\ell(Y_{tk,ij}\mid u_i)]+\log p(u_i^{\ast})-\log p(u_i)\}$.
Receiver effects $v_j$, $j\neq j_v$, are updated analogously, summing over senders $i\neq j$.
Since these effects are shared by all slices, the ratio involves every observed network, not only those of a given regime.

\paragraph{Update of the transition parameters $\boldsymbol{\Gamma}$ and $\boldsymbol{\Lambda}$.}
First, the shift along the non-identified ridge of Section~\ref{sec:hmm} is drawn from its full conditional, $
c_g\sim\mathcal{N}\!\left(m_g,\;\kappa^{-1}\right)$, $
\kappa=\frac{G}{\sigma_\gamma^2}+\frac{K}{\sigma_\lambda^2}$, $
m_g=\kappa^{-1}\Bigl(\sigma_\lambda^{-2}\sum_{k}\lambda_{kg}-\sigma_\gamma^{-2}\sum_{h}\gamma_{hg}\Bigr)$,
and applied as $\gamma_{hg}\leftarrow\gamma_{hg}+c_g$, $\lambda_{kg}\leftarrow\lambda_{kg}-c_g$, for $g<G$; like the translation of $\mathbf{Z}_g$, this step involves only the priors and leaves the posterior invariant, whereas normalising to the canonical representative of Equation~\eqref{eq:gamma_delta_identified} would change the prior density without an acceptance step.
Each free $\gamma_{hg}$, $g<G$, is then updated with proposal $\gamma_{hg}^{\ast} = \gamma_{hg} + \epsilon$, $\epsilon \sim \mathcal{N}(0, \tau_\gamma^2)$, accepted with probability $\min(1, R_\gamma)$, where $R_\gamma = \exp\{ \sum_{k=1}^K \sum_{t=2}^T [\log p_{S_{t-1,k}, S_{tk}}^{\ast(k)} - \log p_{S_{t-1,k}, S_{tk}}^{(k)}] + \log p(\gamma_{hg}^{\ast})-\log p(\gamma_{hg}) \}$.
As the softmax acts row-wise, a change in $\gamma_{hg}$ modifies all transitions out of regime $h$, which all enter the ratio; $\lambda_{kg}$ is updated analogously, using only the transitions of layer $k$.

\subsection{Forward filtering backward sampling}
\label{sec:ffbs}
Instead of evaluating all $G^T$ possible paths, FFBS \citep{carter1994gibbs,fruhwirth1994data} samples the regime sequence $\{S_{tk}\}_{t=1}^T$ of layer $k$ jointly from its full conditional, which makes it a Gibbs step.
For fixed layer $k$ and current values of all other parameters, define the forward message $a_t(g)=p(\mathbf{Y}_{1k},\ldots,\mathbf{Y}_{tk},S_{tk}=g)$, $g=1,\ldots,G$.

\paragraph{Forward filtering and backward sampling.}
For numerical stability the recursion is computed on the logarithmic scale.
At $t=1$, $\log a_1(g) = \ell(\mathbf{Y}_{1k} \mid g) + \log \omega_g$, where $\ell(\mathbf{Y}_{tk}\mid g)$ is the log-likelihood of slice $(t,k)$ under regime $g$ at the current parameter values, and for $t = 2, \ldots, T$, $\log a_t(g) = \ell(\mathbf{Y}_{tk} \mid g) + \log(\sum_{h=1}^G \exp\{\log a_{t-1}(h) + \log p_{hg}^{(k)}\})$, with the inner sum computed by log-sum-exp.
Slices with no observations contribute $\ell(\mathbf{Y}_{tk}\mid g)=0$ for all $g$, so their states are informed by the transition model only.

The final state is drawn from $P(S_{Tk} = g \mid \mathbf{Y}_{1k}, \ldots, \mathbf{Y}_{Tk}) \propto a_T(g)$ and, for $t=T-1,\ldots,1$, each state from $P(S_{tk} = g \mid S_{t+1,k}, \mathbf{Y}_{1k}, \ldots, \mathbf{Y}_{Tk}) \propto a_t(g) \, p_{g, S_{t+1,k}}^{(k)}$, with probabilities normalised over $g=1,\ldots,G$.
The $K$ regime sequences are conditionally independent and sampled separately.
The recursion costs $O(TG^2)$ per layer, but the dominant cost is the evaluation of the $KTG$ slice log-likelihoods, each of order $n^2$, giving $O(KTGn^2)$ per iteration.

\subsection{Label switching correction}
\label{sec:relabeling}
To align the draws (Section~\ref{sec:hmm}), we apply the Equivalence Classes Representatives (ECR) algorithm \citep{papastamoulis2010artificial} to the regime sequences $\{S_{tk}^{(m)}\}$.
The reference labelling is the modal regime sequence of a reference chain, $S_{tk}^{\mathrm{ref}}=\operatorname{mode}\{S_{tk}^{(m)}\}$, and for each retained draw $m$ ECR selects the permutation $\sigma^{(m)}$ that maximises the agreement between $\{\sigma^{(m)}(S_{tk}^{(m)})\}$ and the reference.
The permutation is applied to every regime-indexed quantity of draw $m$: the regime sequences, $\{\mathbf{Z}_g^{(m)}\}$, the rows and columns of $\boldsymbol{\Gamma}^{(m)}$ and the columns of $\boldsymbol{\Lambda}^{(m)}$; since $\boldsymbol{\Gamma}$ and $\boldsymbol{\Lambda}$ are not identified, we report only the transition probabilities $\mathbf{P}^{(k)}$, computed draw by draw from the permuted parameters.
The sender and receiver effects do not depend on the labels.
In both applications ECR selects the identity permutation within each chain, so relabelling only aligns different chains; in the simulation study, the estimated regimes are matched to the true ones afterwards, to compute the recovery metrics.
ECR does not remove the rotational and reflective invariance of the configurations (Section~\ref{sec:lsm}), which does not affect distances and the quantities derived from them, computed on the unaligned draws.
For graphical display, each centred draw of $\mathbf{Z}_g$ is aligned by orthogonal Procrustes \citep[][Ch.~7]{hoff2002latent, dryden2016statistical} to the mean of the aligned draws of the same regime, after relabelling and using only rotations and reflections, since the likelihood is not invariant to scaling.
Each slice $(t,k)$ is summarised by its modal regime $\hat{S}_{tk} = \operatorname{mode}\{S_{tk}^{(m)}\}_{m=1}^M$ and by the posterior probability $\hat{p}_{tk} = \max_{g} M^{-1}\sum_{m=1}^{M} \mathbbm{1}(S_{tk}^{(m)} = g)$, computed from the $M$ retained, relabelled draws; values close to one indicate an essentially certain assignment.

\subsection{Model selection}
\label{sec:model_selection}
The number of regimes $G$ is chosen among candidate values mainly by the Deviance Information Criterion (DIC) \citep*{spiegelhalter2002bayesian}.

\paragraph{Deviance Information Criterion.}
Let $D(\Theta) = -2 \sum_{t=1}^{T}\sum_{k=1}^{K}\ell(\mathbf{Y}_{tk} \mid S_{tk}, \mathbf{Z}_{S_{tk}},\alpha_{tk},\beta_{tk},r_{tk})$ denote the deviance at $\Theta$.
The DIC is $\mathrm{DIC} = \bar{D} + p_D$, where $\bar{D} = \mathbb{E}[D(\Theta)\mid\mathcal{Y}]$ is the posterior mean deviance and $p_D = \bar{D}-D(\bar{\Theta})$ the effective number of parameters.
The plug-in value $\bar{\Theta}$ combines the modal assignments $\hat S_{tk}$ with the post-processed posterior means of $\{\mathbf{Z}_g\}$, $\{\alpha_{tk}\}$ and $\{\beta_{tk}\}$, and with $\bar r_{tk}=\bar a_{tk}^{-2}$, where $\bar a_{tk}$ is the posterior mean of $a_{tk}$, since the posterior mean of $r_{tk}$ may not exist under the prior of Section~\ref{sec:emission_hnb}.
Under the sender--receiver extension, the deviance uses Equation~\eqref{eq:mu_sr} and $\bar{\Theta}$ includes the posterior means of the sender and receiver effects.
As the deviance is conditional on the regime sequences, this is a conditional DIC in the terminology of \citet*{celeux2006deviance}: it does not involve the transition parameters and, as known for missing-data models, can yield $p_D<0$, so we use it only to compare values of $G$.

\paragraph{Selection procedure.}
For each candidate $G$, we run several chains with different seeds, summarise each criterion across chains and exclude values with $p_D<0$.
In the real-data applications, $G$ is retained only if the minimum regime occupancy of the modal regime sequence exceeds 10\% and the mean pairwise agreement between the modal sequences of different chains (the proportion of slices assigned to the same regime once labels are aligned) exceeds 0.70: with the true number of regimes unknown, a $G$ for which chains find different regime sequences does not identify a stable set of regimes, even when its DIC is the lowest.
Among the retained values, we select the one with the lowest DIC averaged across chains; when the base and sender--receiver specifications select different values of $G$, we report both and adopt the specification with the lower DIC at its own selected $G$.
In the simulation study of Section~\ref{sec:simulation}, where the true number of regimes is known, fits with $p_D<0$ or a minimum occupancy below 5\% are excluded and no agreement is required; the lower threshold avoids excluding configurations valid by construction, such as the rare-regime scenario.

\paragraph{Regime-separation diagnostic.}
The diagnostic $\widehat\Delta_{\min}(G)$ of Equation~\eqref{eq:delta_min_def} is also part of the selection rule: reported with the minimum regime occupancy for the selected model, it assesses whether the inferred regimes are genuinely distinct, subject to the caveats of Section~\ref{sec:separation}; its empirical behaviour is studied in Section~\ref{sec:simulation}.

\section{Simulation study}
\label{sec:simulation}
The simulation study assesses the recovery of regimes, latent configurations and parameters as regime separation decreases. Robustness to network sparsity, count dispersion up to the Poisson limit, transition persistence, network dimensions and two misspecifications (latent dimension and number of fitted regimes) is also assessed. Further, we study regime occupancy and $\widehat\Delta_{\min}$ in a null configuration with no regime structure, in over-specified fits and in the case of a rare regime.

\subsection{Design}
\label{sec:sim_design}
Simulated scenarios vary one factor at a time around a common baseline (Appendix Table~S1), which keeps the computational load manageable and avoids combinations with too little data for the number of parameters, such as a large $G$ with a short time series.

\paragraph{Baseline configuration.}
Unless otherwise stated, each dataset has $n=40$ nodes, $T=20$ time points, $K=3$ layers, $G_{\mathrm{true}}=3$ latent regimes and latent dimension $d=2$, with $\pi_{tk}=0.30$, $r_{tk}=1$ and $\alpha_{tk}=2.0$ for all slices, transition persistence $\rho=0.80$ and maximally separated configurations.

\paragraph{Data generation and prior settings.}
The latent positions are not drawn from their Gaussian prior, a setting already examined in \citet{manente2026dynamiclatentspacemodel}, but from a clustered layout that controls regime separation directly: in each regime, nodes are allocated to three clusters whose centres are equally spaced on a circle of radius $2$, each position is the centre of its cluster plus independent $\mathcal{N}(0,0.35^2)$ noise in each coordinate, and the configurations are then centred.
Each node stays in its baseline cluster in each regime with probability $\mathrm{overlap}$ and is otherwise reassigned uniformly at random, independently across regimes, so only the prior on the latent positions is slightly misspecified.
In the baseline transition matrix every regime has persistence $\rho$, with the remaining probability split evenly among the other regimes; the layer deviations are drawn as $\lambda_{kg}\sim\mathcal{N}(0,0.5^2)$ and centred across layers (with uniform initial distribution).
Unless otherwise stated, the model is fitted with $G=G_{\mathrm{true}}$, $d=2$ and the priors of Appendix Table~S10.

\paragraph{Regime separation.}
The generator parameter $\mathrm{overlap}\in[0,1]$ ranges from independently generated configurations ($\mathrm{overlap}=0$) to configurations sharing the baseline partition and differing only through the positional noise ($\mathrm{overlap}=1$, which gives $\Delta_{\min}=0.015$).
As its effect on the geometry perceived by the likelihood is non-linear, we report results against the realised separation $\Delta_{\min}=\min_{g\neq h}\Delta_{gh}$ (Equation~\eqref{eq:delta}), computed exactly from the generating parameters and normalised by $n(n-1)$; unlike a Procrustes-based similarity, it measures how geometric differences change the distribution of the data (Proposition~\ref{cor:delta_pi}).

\paragraph{Scenario 1.}
In Scenario~1, starting from the baseline, which coincides with the $\mathrm{overlap}=0$ cell of the separation grid, we vary one at a time the regime separation; the sparsity, through $\pi_{tk}$, which by Proposition~\ref{cor:delta_pi} changes the information available to distinguish regimes; the dispersion, from strongly overdispersed counts to the Poisson limit, where $r_{tk}$ is only weakly identified; the transition persistence $\rho$, which sets how often regime changes are observed; and $n$, $T$ and $K$ (levels in Appendix Table~S1).
Three further configurations change the regime structure: data generated with $G_{\mathrm{true}}=5$; a rare regime, whose stationary occupancy is made substantially lower than that of the other two by reducing the utility of moving into regime~1, to check whether the model detects it or absorbs it into the others; and a null configuration, in which all regimes share the same configuration ($\mathrm{overlap}=1$, no positional noise, so that $\Delta_{\min}=0$) and a model fitted with $G=3$ has no regime structure to recover.

\paragraph{Scenario 2: misspecification.}
In Scenario~2, positions are generated in $d=3$ dimensions and the model is fitted with $d=2$, a mild misspecification because the cluster centres lie in the first two dimensions and the third coordinate carries only positional noise; and the baseline datasets, generated with $G_{\mathrm{true}}=3$, are refitted with $G=2$, which forces two true regimes to merge, and with $G=4$ and $G=5$, which leave one or two regimes more than the data support, so that the fits can be compared replicate by replicate.

\paragraph{Overall design and inference.}
Scenario~1 has $29$ configurations and Scenario~2 adds one with newly generated data, which with $20$ replicates each gives $600$ simulated datasets.
Every dataset is fitted with $G=G_{\mathrm{true}}$, and the baseline, rare-regime and null datasets also with $G=2$, $4$ and $5$, each fit using $4$ independent MCMC chains of $20{,}000$ iterations with $10{,}000$ of burn-in.

\subsection{Results}
\label{sec:sim_results}
Full results for every configuration are in the Appendix; Table~\ref{tab:sim_summary} collects the headline numbers for five representative configurations of Scenario~1.

\begin{figure}[t]
\centering
\includegraphics[width=\textwidth]{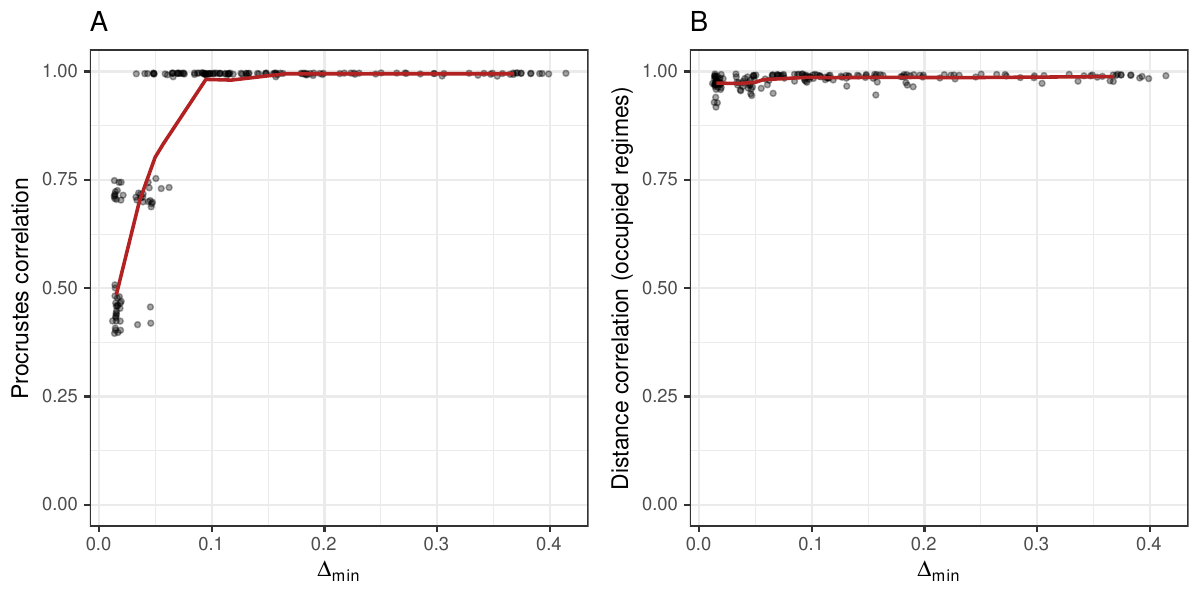}
\caption{Simulation study: latent-space recovery against $\Delta_{\min}$. A: Procrustes correlation with the true configurations, averaged over regimes; B: correlation with the true pairwise distances, minimum over occupied regimes. Points: replicates; line: mean.}
\label{fig:sim_latent}
\end{figure}

\paragraph{Latent geometry.}
The two measures of latent-space recovery in Figure~\ref{fig:sim_latent} diverge as overlap increases.
When regimes are nearly indistinguishable, the model concentrates the slices on a single regime (Table~\ref{tab:sim_summary}, last column) and the geometry of the empty regimes is driven by the prior, which lowers the Procrustes correlation, while the distance correlation stays above $0.97$ for every occupied regime across the whole range of separations: the model does not build a geometry for regimes that the data do not support.
Panel~B is, however, a conditional measure: in the null configuration the distance correlation is $0.988$ even though there is no regime structure, because the single occupied regime has a geometry compatible with the undifferentiated data, so the two panels should be read together with regime occupancy.

\begin{figure}[t]
\centering
\includegraphics[width=\textwidth]{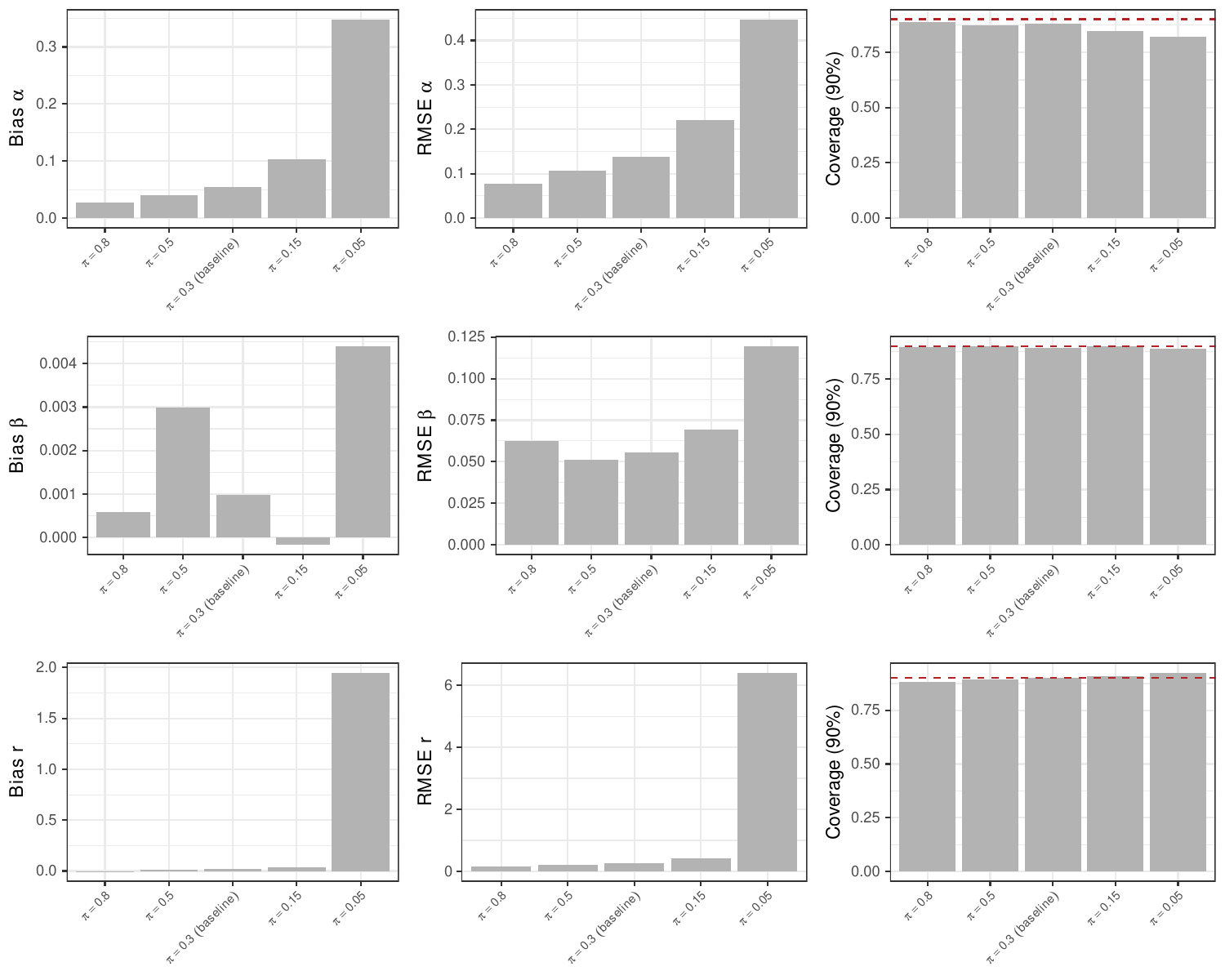}
\caption{Simulation study: bias, RMSE and coverage (columns; dashed line: nominal $90\%$) of $\alpha_{tk}$ (coverage for $\tilde\alpha_{tk}$), $\beta_{tk}$ and $r_{tk}$ (rows) against decreasing $\pi_{tk}$. Full grid in Appendix Table~S2.}
\label{fig:sim_params}
\end{figure}

\paragraph{Network-specific parameters.}
Bias and RMSE for the intercept and for $r_{tk}$, informed only by the positive edges, grow as $\pi_{tk}$ falls from $0.80$ to $0.05$ (Figure~\ref{fig:sim_params}), whereas $\beta_{tk}$ is essentially unaffected, with bias close to zero and coverage between $89\%$ and $90\%$.
Coverage of $\tilde\alpha_{tk}$ is slightly below the nominal $90\%$ level, between $82\%$ and $89\%$ across sparsity levels (Table~\ref{tab:sim_summary}) and between $83\%$ and $89\%$ across separation levels, while coverage of the raw intercept $\alpha_{tk}$ falls from $83$--$84\%$ for well-separated regimes to $38\%$ at $\Delta_{\min}=0.015$, which supports reporting the identified contrast (Section~\ref{sec:emission}).
As predicted by Proposition~\ref{cor:delta_pi}, sparsity reduces separation: from the baseline to $\pi_{tk}=0.05$, the true $\Delta_{\min}$ falls from $0.368$ to $0.061$, equal to the predicted value $0.368\times0.05/0.30$, and the empirical $\widehat\Delta_{\min}$ from $0.344$ to $0.031$.
Regime recovery is nevertheless almost unaffected (ARI $0.991$), although the latent geometry is recovered less accurately (distance correlation $0.890$, the lowest in the design): even very sparse networks carry enough information to classify slices when the configurations are well separated.
The dispersion parameter is the hardest to recover when overdispersion is weak: bias and RMSE for $r_{tk}$ are $1.42$ and $5.05$ at $r_{tk}=5$, $49.0$ and $69.7$ at $r_{tk}=20$, and $-57.5$ and $80.3$ at $r_{tk}=200$ (Appendix Table~S2), as expected since the likelihood becomes flat in $r_{tk}$ near the Poisson limit.
The chains nevertheless agree on $r_{tk}$, with $\widehat R$ below $1.02$ for $99\%$ of the slices (Appendix Table~S7), and the regimes are recovered as well as elsewhere (ARI $=1$ at all three levels).

\begin{figure}[t]
\centering
\includegraphics[width=\textwidth]{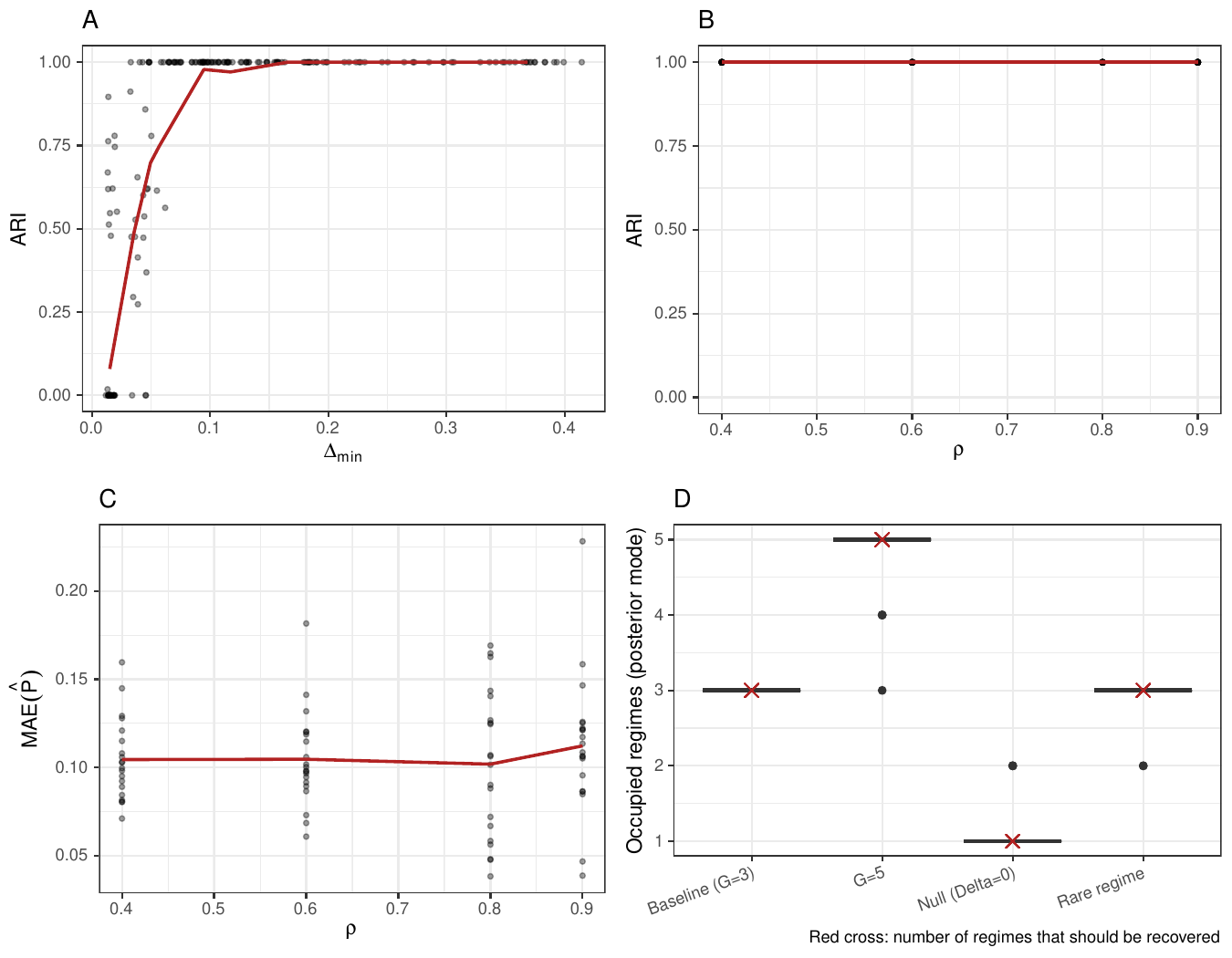}
\caption{Simulation study: regime recovery. ARI against $\Delta_{\min}$ (A) and $\rho$ (B); C: mean absolute error of the transition matrix against $\rho$; D: occupied regimes at baseline, with $G_{\mathrm{true}}=5$, with a rare regime and in the null configuration (red cross: true value).}
\label{fig:sim_hmm}
\end{figure}

\paragraph{Regime dynamics.}
ARI decreases with separation (Figure~\ref{fig:sim_hmm}, Panel~A): it is $1.00$ at maximal separation, $0.747$, $0.698$ and $0.490$ at $\mathrm{overlap}=0.94$, $0.955$ and $0.97$ ($\Delta_{\min}=0.057$, $0.050$ and $0.036$), and $0.079$ at $\mathrm{overlap}=1$ ($\Delta_{\min}=0.015$).
Transition persistence leaves regime recovery unaffected, with ARI equal to $1$ for all values of $\rho$ (Panel~B), but affects the estimated transition matrix (Panel~C): its mean absolute error rises from $0.104$ at $\rho=0.40$ to $0.112$ at $\rho=0.90$ and its coverage falls from $90\%$ to $84\%$, since with $T=20$ a highly persistent chain produces few observed transitions and hence little information on the off-diagonal probabilities.
On average the model occupies $4.75$ regimes when $G_{\mathrm{true}}=5$, $2.90$ when one of three regimes is rare, and $1.15$ in the null configuration (Panel~D).
With $T=20$ the realised share of the rare regime varies across replicates, with a mean of $10\%$: in two replicates the simulated chains never visit it, and in the remaining $18$ the model detects it every time (ARI $=1$).

\paragraph{Misspecification (Scenario~2).}
\label{sec:sim_scen2}
With the latent dimension underspecified ($d=3$ fitted with $d=2$), ARI is $1$ and the distance correlation $0.980$ (Appendix Table~S9a); the cost of this mild misspecification appears only in the coverage of the latent coordinates, which falls to $0.566$ (Appendix Table~S3).
With a wrong number of regimes (Appendix Table~S9b), DIC is higher for $G=2$ than for $G=3$ in every replicate, but differs by only about two units between $G=3$, $4$ and $5$, with $G=3$ lowest in only $6$ of the $20$ replicates.
The fits with $G=4$ and $G=5$ leave the additional regimes empty, occupying exactly three regimes in every replicate, with the same partition as the $G=3$ fit (ARI $=1$) and the same $\widehat\Delta_{\min}$.
The rare-regime and null datasets behave alike; in the null configuration, the fits with $G=3$, $4$ and $5$ occupy a single regime in $17$ to $20$ of the $20$ replicates.
DIC and $\widehat\Delta_{\min}$ therefore do not detect over-specification, whereas regime occupancy does, which motivates the occupancy requirement of Section~\ref{sec:model_selection}.

\begin{figure}[t]
\centering
\includegraphics[width=.7\textwidth]{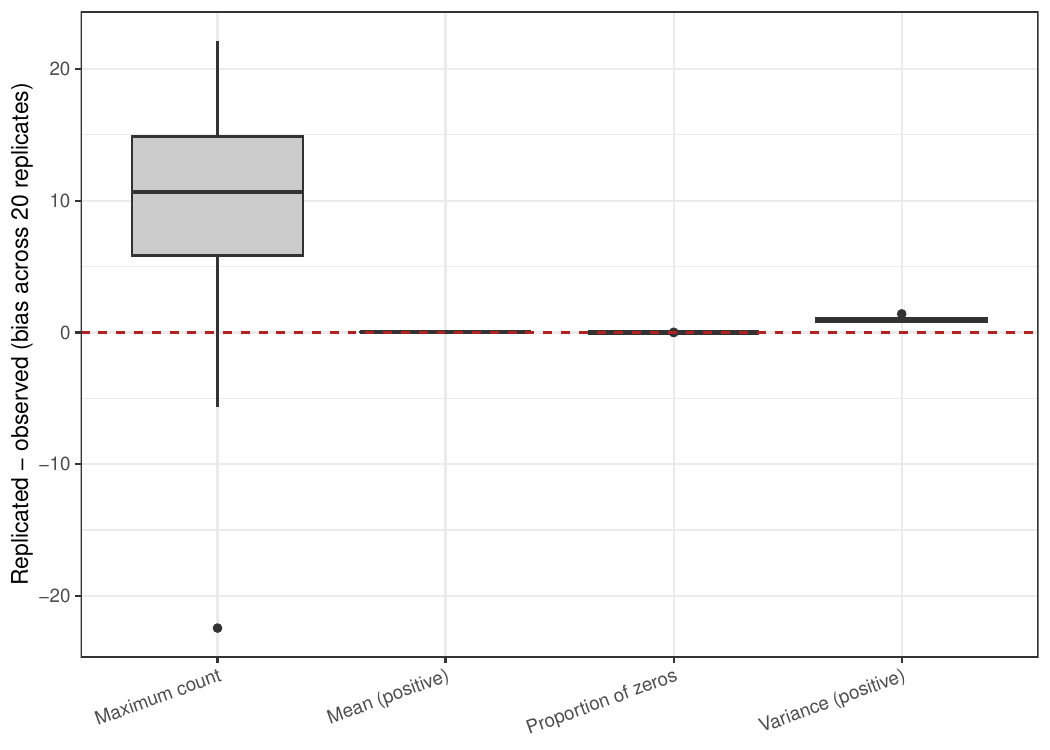}
\caption{Simulation study: posterior predictive checks at baseline. Mean replicated minus observed statistic over the $20$ replicates (mean and variance of positive counts); dashed line: zero.}
\label{fig:sim_ppc}
\end{figure}

\paragraph{Posterior predictive checks.}
Posterior predictive checks use $1{,}000$ draws from the first chain of each replicate (Figure~\ref{fig:sim_ppc}): coverage of the $5$--$95\%$ posterior predictive interval is $100\%$ for both the proportion of zero edges and the mean of the positive counts, and $35\%$ for their variance, which varies across the design (Appendix Table~S6).

\paragraph{Convergence.}
Convergence is assessed through the rank-normalised split-$\widehat R$ of \citet*{vehtari2021rank}, computed between the chains for every model parameter (Appendix Table~S7): across the design, the median $\widehat R$ is $1.00$ for every type of parameter, and $\beta_{tk}$ never exceeds $1.1$.
At $\mathrm{overlap}=0.955$ and $0.97$, where ARI decreases most (Figure~\ref{fig:sim_hmm}, Panel~A), the chains occasionally disagree on the regime sequence, and between $6\%$ and $15\%$ of the intercepts, dispersions, latent distances and transition probabilities have $\widehat R>1.1$.
Elsewhere the chains agree on the regimes; where they do not fully converge, as near the Poisson limit and in the null configuration, they differ in the latent distances ($5$--$20\%$ above $1.1$) rather than in the regime sequence.

\paragraph{Regime-separation diagnostic.}
In the rare-regime configuration, $\widehat\Delta_{\min}=0.33$ is close to the true separation ($0.37$), because the rare regime is occupied whenever it is present.
In the null configuration ($0.28$ against a true value of $0$) and at $\mathrm{overlap}=1$ ($0.21$ against $0.015$), the diagnostic lies far above the true value because one or more fitted regimes are nearly empty (Table~\ref{tab:sim_summary}), the inflation anticipated in Section~\ref{sec:separation}.

\begin{table}[t]
\centering
\caption{Simulation study, Scenario~1: summary. ARI: adjusted Rand index (undefined for the null configuration); Dist.\ corr.\ (occ.): correlation between estimated and true latent distances, minimum over occupied regimes; Coverage: of $90\%$ credible intervals for $\tilde\alpha_{tk}$ (Equation~\eqref{eq:alpha_contrast}); Occ.\ regimes: mean number of occupied regimes out of three.}
\label{tab:sim_summary}
\setlength{\tabcolsep}{4pt}
\begin{tabular}{p{3cm}rrrrr}
\toprule
Configuration & ARI & \shortstack[r]{Dist.\ corr.\\(occ.)} & Coverage & $\widehat\Delta_{\min}$ & \shortstack[r]{Occ.\\regimes} \\
\midrule
Baseline ($\Delta_{\min}=0.368$) & 1.000 & 0.988 & 0.882 & 0.344 & 3.00 \\
Overlap $=1$ ($\Delta_{\min}=0.015$) & 0.079 & 0.973 & 0.892 & 0.208 & 1.15 \\
Sparsest ($\pi_{tk}=0.05$) & 0.991 & 0.890 & 0.823 & 0.031 & 3.00 \\
Rare regime ($\Delta_{\min}=0.370$) & 1.000 & 0.973 & 0.897 & 0.328 & 2.90 \\
Null ($\Delta_{\min}=0$) & -- & 0.988 & 0.875 & 0.279 & 1.15 \\
\bottomrule
\end{tabular}
\end{table}

\section{Real data applications}
\label{sec:applications}
In both applications we fix $d=2$, which allows direct graphical representation of the latent space (the simulation study indicates that regime recovery is robust to a non-optimal dimension), run eight chains for each candidate $G$, select $G$ as described in Section~\ref{sec:model_selection}, and set the prior hyperparameters on the scale of the observed counts (Appendix Table~S10).

\subsection{GDELT: geopolitical events}
\label{sec:gdelt}
The GDELT 2.0 Event Database records machine-coded international events extracted from worldwide news coverage and classified with the CAMEO taxonomy \citep{leetaru2013gdelt}.
From the public BigQuery table \texttt{gdelt-bq.gdeltv2.events} we retain root events (\texttt{IsRootEvent = 1}) and aggregate them into monthly country-to-country counts, with the four QuadClasses (verbal cooperation, material cooperation, verbal conflict, material conflict) as layers.
The nodes are the $n=45$ countries with the largest total event volume over the complete calendar years 2015--2024, which coincide with those selected over the whole observation period, and the panel covers $T=138$ months, from March 2015 to August 2026 ($KT=552$ slices, approximately $57.4$ million events).
Verbal cooperation is markedly denser and more voluminous than the other three layers (Appendix Table~S11).
We fit the base and sender--receiver (SR) specifications for $G=2,\ldots,8$, with $30{,}000$ iterations and $10{,}000$ of burn-in.
DIC decreases with $G$ throughout for both, so the choice rests on the occupancy and agreement constraints: values $G\geq4$ leave at least one regime (nearly) empty, and $G=3$ is selected for both (Appendix Table~S12).
At $G=3$ the SR model lowers DIC by about $333{,}000$ and is used below; six of the eight chains agree on the same solution, and we report the chain with the highest agreement with the others.

\begin{figure}[t]
\centering
\includegraphics[width=\textwidth]{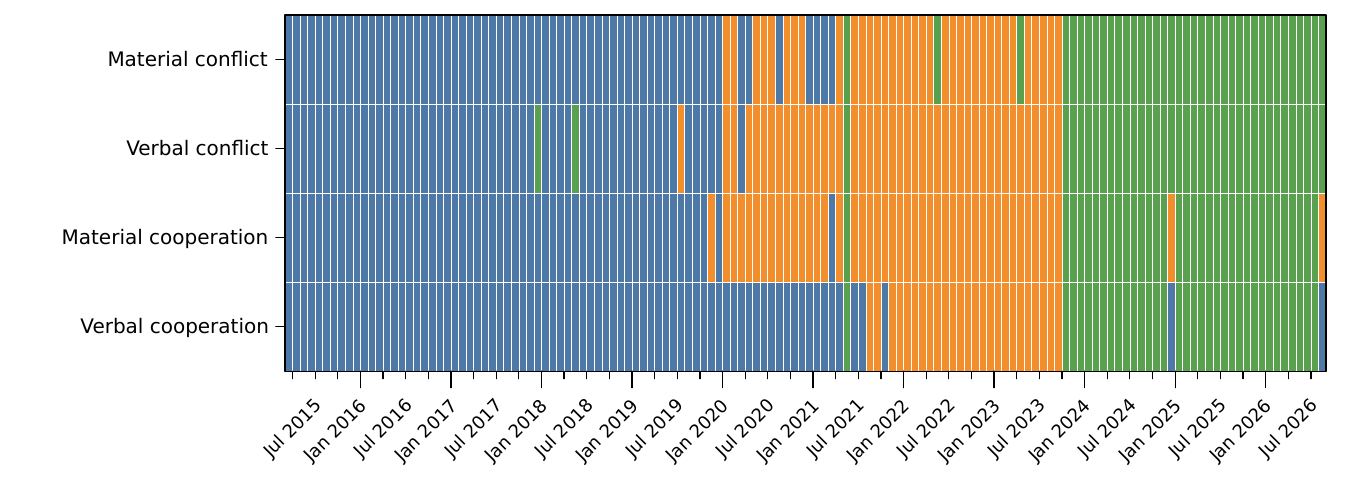}
\caption{GDELT, sender--receiver model, $G=3$: modal regime by month and layer (blue, orange, green: regimes~1, 2, 3).}
\label{fig:gdelt_regimes}
\end{figure}

\paragraph{Regime dynamics.}
Until 2019 the four layers are essentially in regime~1 (Figure~\ref{fig:gdelt_regimes}). Then, they move to regime~2 at different times: material cooperation in January 2020, verbal conflict in April 2020, material conflict in June 2021 and verbal cooperation in November 2021.
Regime~2 persists until September 2023, and in October 2023 all four layers move to regime~3 in the same month, a simultaneous switch that the model does not impose (Section~\ref{sec:hmm}) and that therefore comes from the data.
The entry into regime~2 coincides with the onset of the COVID-19 pandemic, which altered both international interactions and the news coverage on which GDELT relies \citep*{hammond2014using,wang2016growing}, and with the killing of Qasem Soleimani and the downing of flight PS752 in January 2020, which generated intense interactions among the United States, Iran, Iraq and Ukraine \citep{aljazeera2020soleimani,aljazeera2020ps752}.
Regime~2 then spans the Russia--Ukraine crisis and the full-scale invasion of February 2022 \citep{cfr2026ukraine}, which does not open a new regime, because the latent positions of Russia and Ukraine already almost coincide in regime~2.
The switch to regime~3 coincides with the Hamas attack of 7 October 2023 and the ensuing regional escalation \citep{aljazeera2023hamas}; the same configuration appears in all four layers in May 2021, during the eleven-day Gaza conflict \citep{ocha2021gaza}, and in the verbal conflict layer in December 2017, so regime~3 is a recurring configuration rather than a late-period effect.

\begin{figure}[t]
\centering
\includegraphics[width=\textwidth]{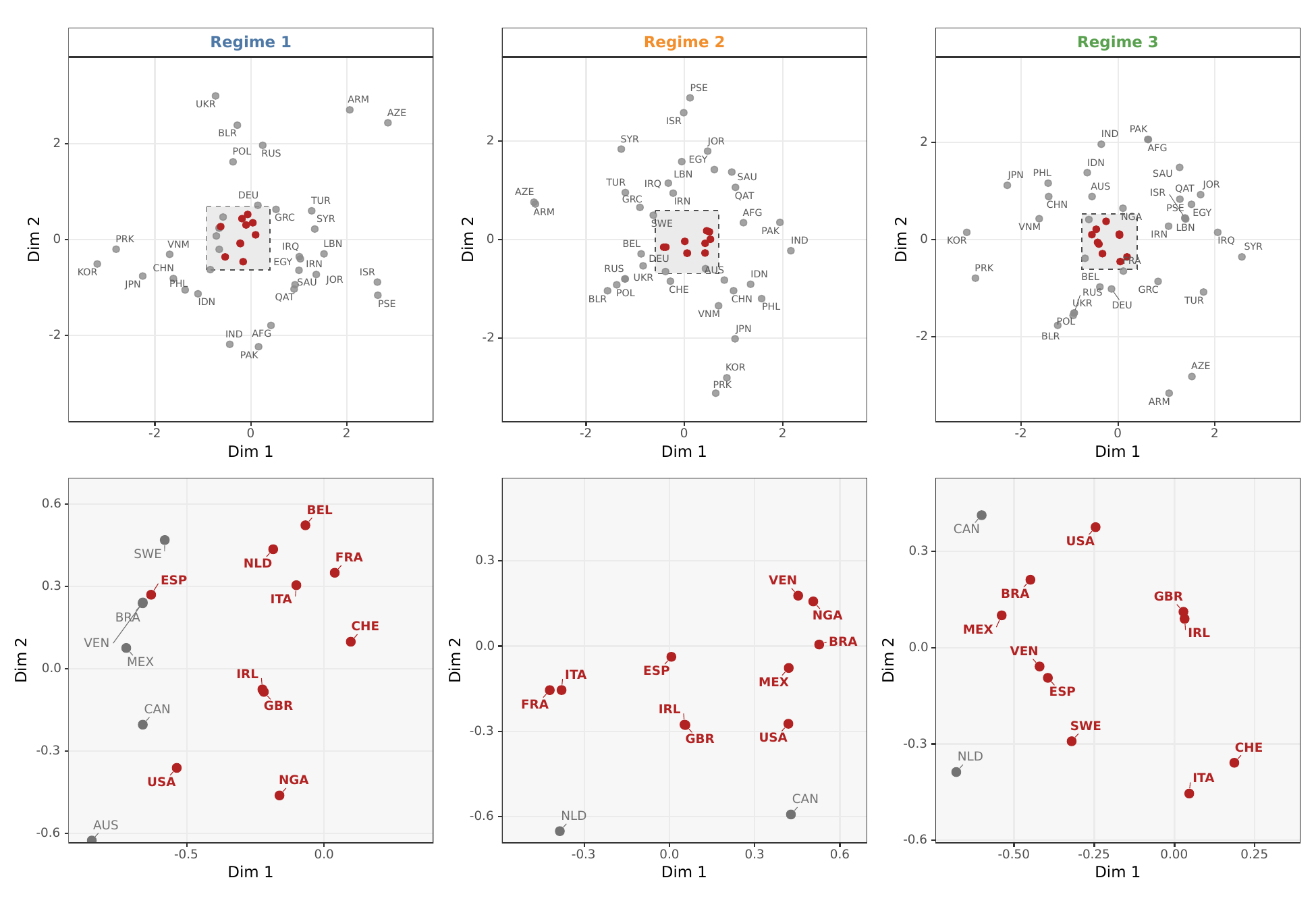}
\caption{GDELT, sender--receiver model, $G=3$: posterior mean latent positions by regime. Red: the ten countries closest on average to all others; bottom row: zoom on the dashed box.}
\label{fig:gdelt_latent}
\end{figure}

\paragraph{Latent geometry.}
The three regimes have similar mean distances ($2.16$, $2.08$ and $2.15$), so a change of regime mainly reallocates the expected volume among dyads rather than scaling it (Figure~\ref{fig:gdelt_latent}).
In regime~1 Turkey and Syria are among the closest pairs ($D_1(\mathrm{TUR},\mathrm{SYR})=0.38$), consistent with the Syrian war and the Turkish operations in northern Syria \citep{phillips2016battle,aljazeera2019peacespring}, while in the other regimes their expected count is 40--50\% lower.
From regime~1 to regime~2 the expected count of the Russia--Ukraine dyad grows by a factor of about $4$ at fixed activity levels, as their latent distance approaches zero.
In regime~3 Russia and Ukraine remain close, and the reallocation involves Israel: Israel and Palestine also reach a distance close to zero, and the expected counts of Israel--Lebanon and Israel--Iran increase by factors of about $2.7$ and $3.6$ relative to regime~2.
The United States keeps Canada, the United Kingdom and Ireland among its five closest partners in every regime: the regimes reallocate expected interaction within specific regional neighbourhoods, consistent with the regional security complexes of \citet{buzan2003regions}, rather than across the whole network.
The sender and receiver effects capture differences in overall activity across countries, and the latent distance differences between specific pairs: Russia, Israel and Ukraine are among the most active countries after the United States, but not among the most central in any regime.
This separation is useful for GDELT, where the number of recorded events also depends on news coverage, which varies across countries and over time \citep{hammond2014using,wang2016growing}.

\subsection{European Parliament: disagreement between states}
\label{sec:ep}
We use roll-call votes of the ninth European Parliament (EP9) from the VoteWatch Europe dataset \citep*{hix2022votewatch}.
The nodes are the $n=27$ member states, observed in $K=4$ policy layers (economy, environment, foreign affairs and social policy) over the $T=33$ months with plenary sessions between July 2019 and June 2022.
On each vote a member state takes the majority position of its MEPs, and $Y_{tk,ij}$ counts the votes on which $i$ voted in favour and $j$ against, so the network records directed disagreement; unlike pairs of MEPs, national majorities avoid counts driven by delegation size (the correlation between a delegation's total count and its size is $0.992$ for pairs of MEPs and $0.225$ for national majorities).
Since expected counts decrease with latent distance, the latent space reads in reverse relative to a cooperation network: member states that are close are expected to disagree more, and a central position indicates disagreement with many delegations rather than a moderate one (data summary in Appendix Table~S13).
As for GDELT, DIC decreases with $G$ for both specifications, and the occupancy and agreement constraints select $G=2$ for the base model and $G=3$ for the sender--receiver model (Appendix Table~S14), which we use since its DIC is lower by about $14{,}000$; seven of the eight chains agree, and we report the chain with the highest agreement with the others.

\begin{figure}[t]
\centering
\includegraphics[width=\textwidth]{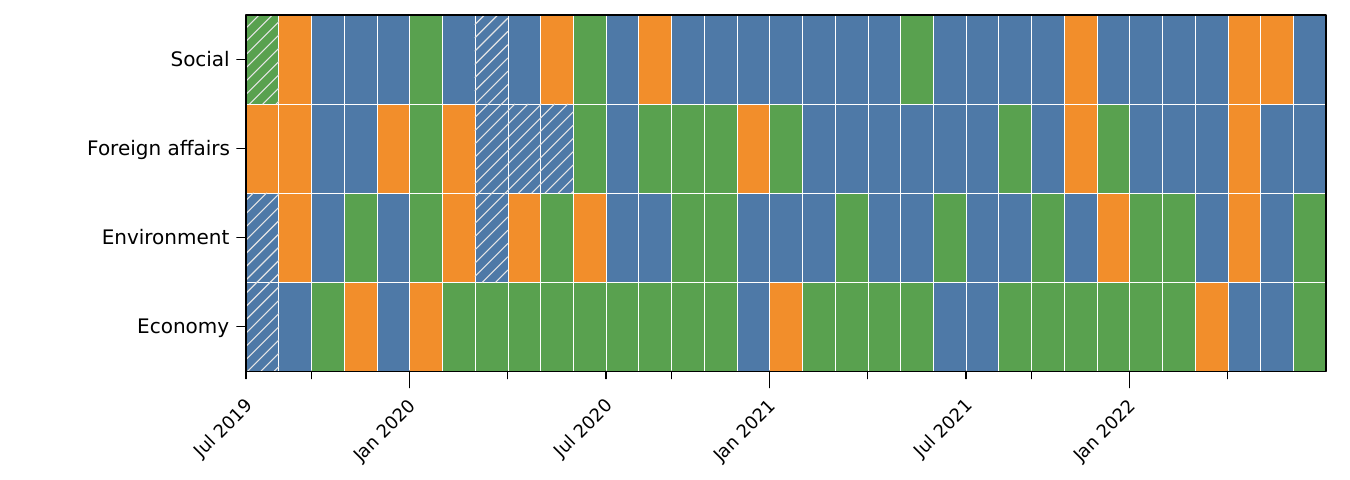}
\caption{European Parliament, sender--receiver model, $G=3$: modal regime by month and layer (blue, orange, green: regimes~1, 2, 3).}
\label{fig:ep_regimes}
\end{figure}

\paragraph{Regime dynamics.}
The layers move repeatedly between the three regimes (Figure~\ref{fig:ep_regimes}), whose slices differ markedly in volume: regime~2 collects low-disagreement months (median $176$ counts per slice, $75\%$ zero dyads), regime~1 intermediate ones ($2{,}494$; $21\%$), and regime~3 high-disagreement months ($6{,}569$; $4\%$), most frequent in the economy layer.
Some transitions coincide with specific parliamentary activity: in the environment layer, regime~3 occurs in October 2020 and June 2022, with divided votes on the European Climate Law and on the revision of the Emissions Trading System \citep{ep2020climatelaw,ep2022ets}, whereas regime~2 appears on three layers in September 2019 and April 2022, when only one Strasbourg plenary session took place \citep{ep2019plenarysep,ep2022plenaryapr}.
The eight layer--month slices without roll-call votes are treated as missing rather than as networks of zero counts (Section~\ref{sec:ffbs}; hatched cells in Figure~\ref{fig:ep_regimes}), so their regime is inferred from the transition dynamics alone; five of them fall between March and May 2020, when the plenary calendar was reduced at the start of the COVID-19 pandemic \citep{ep2020calendar}, so regime~2 also identifies months of reduced parliamentary activity.

\begin{figure}[t]
\centering
\includegraphics[width=\textwidth]{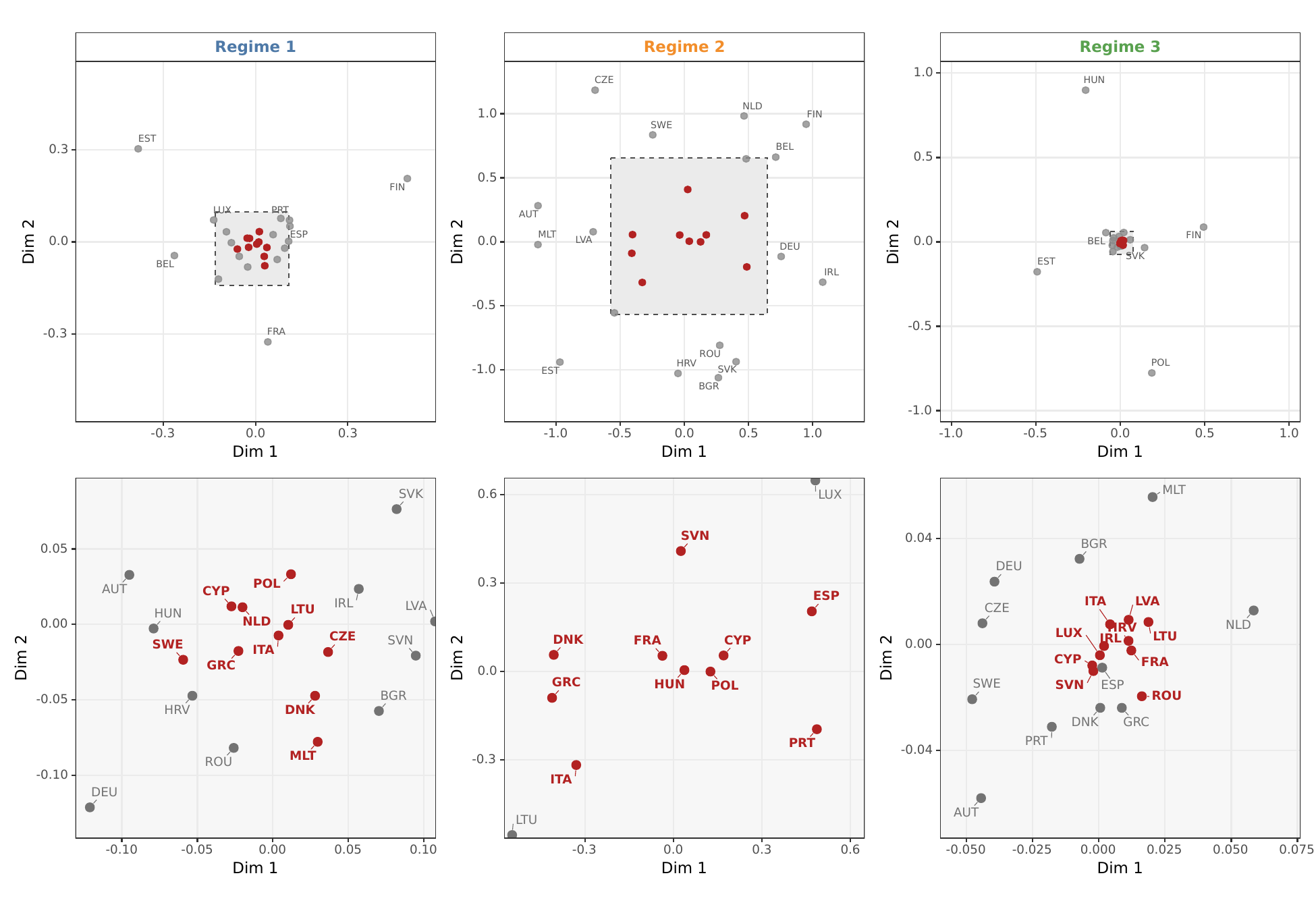}
\caption{European Parliament, sender--receiver model, $G=3$: posterior mean latent positions by regime (close member states are expected to disagree more). Red: the ten states closest on average to all others; purple: Hungary and Poland. Top row: labels outside the dashed box; bottom row: zoom on the box.}
\label{fig:ep_latent}
\end{figure}

\paragraph{Latent geometry.}
The mean latent distance is $0.27$ in regime~1, $1.22$ in regime~2 and $0.28$ in regime~3: the more dispersed configuration of regime~2 reduces the expected disagreement of a typical pair, consistently with the low volume of its slices, while regimes~1 and~3 have similar mean distances but differ in which pairs are close (Figure~\ref{fig:ep_latent}).
Sender--receiver effects and latent distances have effects of comparable size on expected disagreement, which decreases by a factor of about $2.2$ from a short to a long distance and increases by a factor of about $2.4$ from weakly to highly active delegations; in regime~2 the sender--receiver effects are similar across delegations, and the latent distances mainly determine which pairs account for the remaining disagreement.
Poland and Hungary have the largest sender and receiver effects, and their receiver effects exceed their sender effects ($v=1.82$ and $1.83$, $u=1.55$ and $1.45$), so they more often oppose measures that others support than the reverse, consistently with the disputes involving their governments over the rule of law \citep{pech2017illiberalism}.
In regime~2 they are among the most central delegations, so their disagreement is spread across many others, whereas in regime~3 they are the two most peripheral: the latent space moves expected disagreement towards other pairs, while their large effects keep their overall activity high.
The sender--receiver effects thus describe how often each delegation is on either side of a disagreement, and the latent space with whom, consistently with evidence that voting in the European Parliament is organised mainly along political rather than national lines, with national delegations diverging on specific issues \citep*{hix2006dimensions,hix2007democratic,faas2003defect}, and with the absence of a clear East--West divide after enlargement \citep{hix2009after}.

\section{Discussion}
\label{sec:discussion}
The LS-HMM yields a dynamic clustering of longitudinal multilayer networks, in which layer-specific hidden Markov chains move each layer among a few recurrent latent configurations, an additive transition model keeps the number of parameters under control as layers grow, and a Hurdle Negative Binomial emission, optionally with sender--receiver effects, accommodates sparse and overdispersed counts.
In the simulation study, regimes and latent configurations are recovered accurately when well separated, and performance declines gradually as their separation, a quantity computable from the model (Section~\ref{sec:separation}), decreases.
In both applications the model identifies a few recurrent configurations with a clear substantive reading: in GDELT the regimes reallocate interaction within regional neighbourhoods, and a configuration centred on Israel recurs several times before becoming persistent in October 2023; in the European Parliament they separate months of intense and reduced disagreement, while the sender and receiver effects distinguish the overall activity of each delegation from the pattern of disagreement between delegations.
Several extensions are possible.
Mixtures of finite mixtures with telescoping sampling \citep*{fruhwirth2021generalized} would infer $G$ jointly with the other parameters rather than over a grid of values, and the latent dimension could also be selected from the data \citep*{gwee2025latent}.
Time-varying covariates in the transition probabilities would relate external events directly to the timing of regime changes, and a layer-specific term on the diagonal of the transition utilities would let layers differ in their persistence.
The latent geometry could also enter the hurdle component, not only the intensity of the interactions, and exposure terms such as the number of votes in a month could be added.
As the hidden Markov structure and the latent space component do not depend on the emission, the Hurdle Negative Binomial can be replaced by other distributions, such as a Bernoulli for binary networks or continuous distributions for weighted ones.
Finally, a regime could be defined by any subset of the emission parameters, constrained to be identical within the regime and chosen according to the research question, rather than by the latent configuration alone; including the intercepts $\alpha$, for instance, would identify periods with a common overall intensity of interaction, alongside or instead of a common latent geometry.
\section*{Data Availability} Synthetic data from the simulation study are generated as described in the paper.
The GDELT 2.0 Event Database is publicly available (see \citet{leetaru2013gdelt}). The European Parliament data are available from VoteWatch \citep{hix2022votewatch}.

\section*{Code Availability} The \texttt{R} and \texttt{C++} code implementing the model is available at \url{https://github.com/ceciliamanente/countLSHMM}.

\bibliography{sn-bibliography}

\clearpage
\section*{Appendix}
\setcounter{section}{0}
\setcounter{table}{0}
\setcounter{figure}{0}
\renewcommand{\thesection}{S\arabic{section}}
\renewcommand{\thetable}{S\arabic{table}}
\renewcommand{\thefigure}{S\arabic{figure}}
\renewcommand{\thealgorithm}{S\arabic{algorithm}}

\section{Model summary}
\label{sec:summary}
\begin{figure}[H]
\centering
\begin{tikzpicture}[>=Latex, every node/.style={font=\small}]

\node[latent] (S)     at (0,0)       {$S_{tk}$};
\node[obs]    (Y)     at (0,-2.3)    {$Y_{tk,ij}$};
\node[latent] (alpha) at (-2.6,-1.1) {$\alpha_{tk}$};
\node[latent] (beta)  at (-2.6,-2.3) {$\beta_{tk}$};
\node[latent] (r)     at (-2.6,-3.5) {$r_{tk}$};
\node[latent] (Zg)    at (3.4,-2.3)  {$\mathbf{Z}_g$};
\node[latent] (Gamma) at (-0.8,2.0)  {$\boldsymbol{\Gamma}$};
\node[latent] (Lambda) at (0.8,2.0)  {$\boldsymbol{\Lambda}$};

\plate[inner sep=5pt]{pY}{(Y)}{$i\neq j$};
\plate[inner sep=8pt]{pTK}{(S)(alpha)(beta)(r)(pY)(pY-caption)}{$t=1,\ldots,T$; $k=1,\ldots,K$};
\plate[inner sep=6pt]{pG}{(Zg)}{$g=1,\ldots,G$};

\edge {Gamma,Lambda} {S};
\edge {alpha,beta,r,Zg} {Y};
\draw[->] (S) -| node[pos=0.3, above, font=\scriptsize] {selects $g = S_{tk}$} (pG.north);

\draw[->] (S) to[out=160, in=200, looseness=6]
node[left, font=\scriptsize] {Markov} (S);

\end{tikzpicture}
\caption{Plate representation of the LS-HMM: latent quantities in white, observed data shaded; sender--receiver effects omitted.}
\label{fig:plate}
\end{figure}
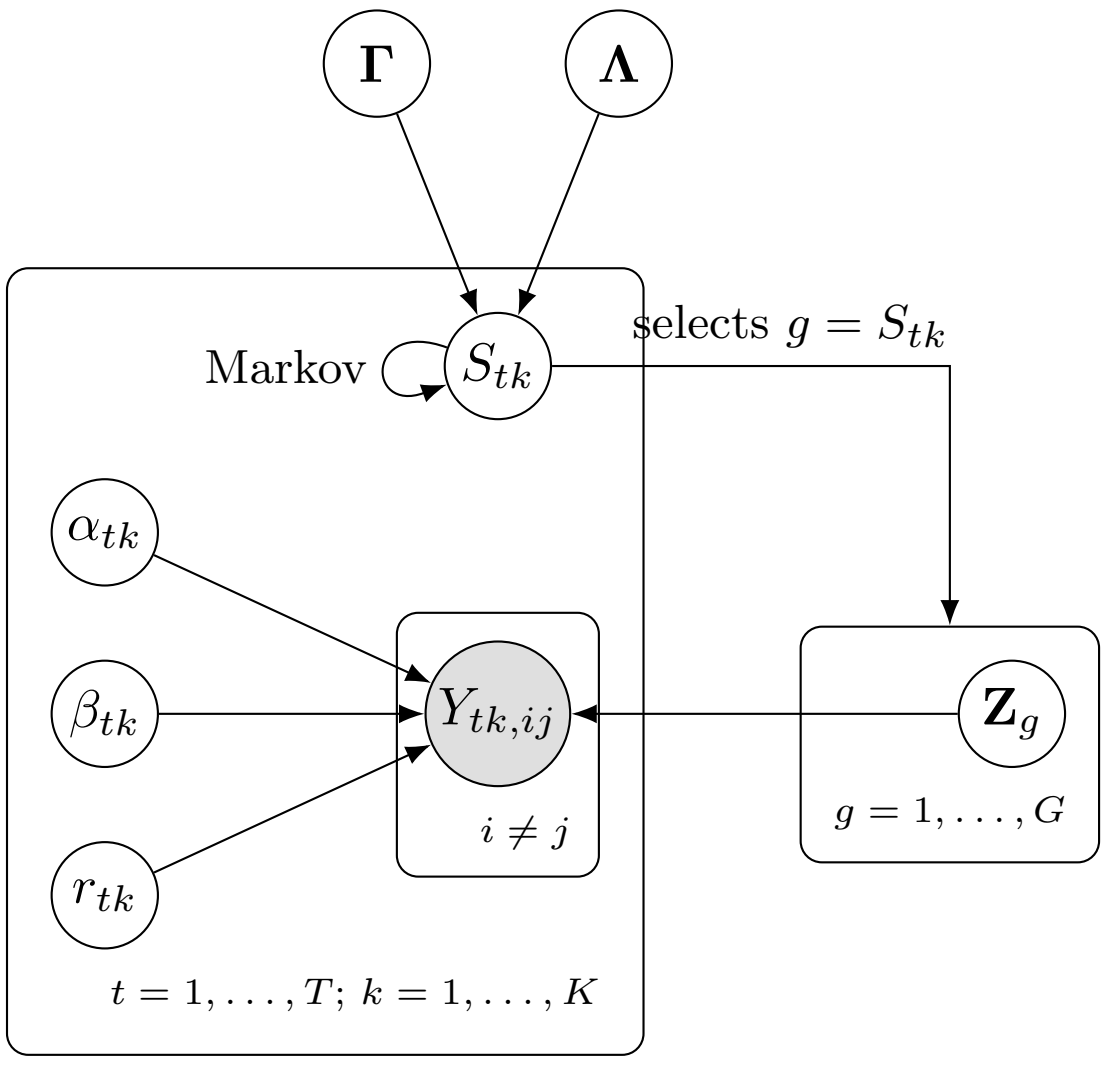

The latent space hidden Markov model (LS-HMM) has a three-level hierarchical structure (Figure~\ref{fig:plate}): the configurations $\{\mathbf{Z}_g\}_{g=1}^{G}$ are the available regimes, the hidden Markov chain $\{S_{tk}\}_{t=1}^T$ of each layer $k$ selects the active configuration at each time point, and, given $S_{tk}=g$, the edge values are generated independently from the HNB emission of Equation~(3) of the main text.
The complete parameter vector, including the regime sequences, is $\Theta=(\{\mathbf{Z}_g\}_{g=1}^G,\allowbreak\{S_{tk}\}_{t,k},\allowbreak\{\alpha_{tk}\}_{t,k},\allowbreak\{\beta_{tk}\}_{t,k},\allowbreak\{r_{tk}\}_{t,k},\allowbreak\boldsymbol{\Gamma},\boldsymbol{\Lambda})$, and becomes $\Theta_{\mathrm{SR}}=\Theta\cup\{u_i\}_{i\neq i_u}\cup\{v_j\}_{j\neq j_v}$ under the sender--receiver extension; posterior inference on $p(\Theta\mid\mathcal{Y})$ uses the algorithm of Section 3.

\section{MCMC algorithm}
\label{sec:Salg}
Algorithm~\ref{alg:mcmc} summarises one run of the sampler described in Section~3.1 of the main text.

\begin{algorithm}[t]
\caption{MCMC scheme for posterior inference in the LS-HMM}
\label{alg:mcmc}
\begin{algorithmic}[1]
\State \textbf{Input:} initial values $\{\mathbf{Z}_g^{(0)}\}$, $\{S_{tk}^{(0)}\}$, $\{\alpha_{tk}^{(0)}, \beta_{tk}^{(0)}, r_{tk}^{(0)}\}$, $\Gamma^{(0)}$, $\Lambda^{(0)}$ (and $\{u_i^{(0)}\}$, $\{v_j^{(0)}\}$ for the SR extension, with $u_{i_u}=v_{j_v}=1$); burn-in length $M_0$ and number of retained draws $M$
\For{$m = 1, \ldots, M_0+M$}
\State \textbf{[1]} For each layer $k$, update $\{S_{tk}\}_{t=1}^T$ jointly by FFBS (Section~3.2 of the main text)
\State \textbf{[2]} For each regime $g$, draw the translation of $\mathbf{Z}_g$ from its full conditional, then update the nodes of $\mathbf{Z}_g$ by MH steps
\State \textbf{[3]} For each slice $(t,k)$, update $\alpha_{tk}$, $\beta_{tk}$, $a_{tk}$ by MH steps, conditionally on $S_{tk}$ and $\mathbf{Z}_{S_{tk}}$
\State \textbf{[3b]} (SR extension only) Update $u_i$, $i\neq i_u$, and $v_j$, $j\neq j_v$, by MH steps
\State \textbf{[4]} Draw the shift along the $(\Gamma,\Lambda)$ ridge from its full conditional, then update $\Gamma$ and $\Lambda$ by MH steps, conditionally on $\{S_{tk}\}$
\State \textbf{[5]} If $m\leq M_0$, adapt the proposal standard deviations 
\EndFor
\State \textbf{Output:} the $M$ post-burn-in draws
\end{algorithmic}
\end{algorithm}

\paragraph{Adaptation.}
During burn-in, the proposal standard deviations $\tau_z$, $\tau_\alpha$, $\tau_\beta$, $\tau_a$, $\tau_\gamma$, $\tau_\lambda$ and, for the sender--receiver extension, $\tau_u$ and $\tau_v$, are adjusted every 100 iterations when the acceptance rate falls outside $[0.20,0.50]$; adaptation then stops, so the retained draws come from a time-homogeneous Markov chain with the posterior as stationary distribution.

\section{Simulation study: complete design}
\label{sec:S1}
Table~\ref{tab:S1} lists the simulation configurations: the groups from \emph{Regime separation} to \emph{Regime structure} form Scenario~1, the last group Scenario~2 (the refits with a wrong number of regimes, which reuse existing datasets, are in Table~\ref{tab:S9b}).
Besides the design values, the table reports the proportion of zero edges per slice (mean over slices) and the variance-to-mean ratio (VMR) of the positive counts per slice (median over slices; $1$ for Poisson counts), averaged over the $20$ replicates.
The proportion of zeros equals $1-\pi_{tk}$ by construction.
The VMR is about $4.5$ at baseline and varies mainly with the dispersion, from $16.0$ at $r_{tk}=0.2$ to $1.9$ at $r_{tk}=200$; it is also higher in the null configuration ($8.7$), where nodes of the same cluster share the same position.

\begin{table}[htbp]
\centering
\caption{Complete simulation design. $n$, $T$, $K$, $G_{\mathrm{true}}$: values fixed by the design. Prop.\ zeros: proportion of zero edges in a slice; VMR: variance-to-mean ratio of the positive counts in a slice (median over slices; $1$ for Poisson counts). Realised values are averaged over $20$ replicates. Groups from \emph{Regime separation} to \emph{Regime structure}: Scenario~1; last group: Scenario~2. The first row ($\mathrm{overlap}=0$) is the baseline configuration.}
\label{tab:S1}
\begin{tabular}{lrrrrrr}
\toprule
Configuration & $n$ & $T$ & $K$ & $G_{\mathrm{true}}$ & Prop.\ zeros & VMR \\
\midrule
\multicolumn{7}{l}{\textit{Regime separation}} \\
$\Delta_{\min}=0.368$ (overlap $=0$)   & 40 & 20 & 3 & 3 & 0.70 & 4.4 \\
$\Delta_{\min}=0.237$ (overlap $=.75$) & 40 & 20 & 3 & 3 & 0.70 & 4.4 \\
$\Delta_{\min}=0.166$ (overlap $=.85$) & 40 & 20 & 3 & 3 & 0.70 & 4.5 \\
$\Delta_{\min}=0.117$ (overlap $=.90$) & 40 & 20 & 3 & 3 & 0.70 & 4.5 \\
$\Delta_{\min}=0.095$ (overlap $=.92$) & 40 & 20 & 3 & 3 & 0.70 & 4.4 \\
$\Delta_{\min}=0.057$ (overlap $=.94$) & 40 & 20 & 3 & 3 & 0.70 & 4.4 \\
$\Delta_{\min}=0.050$ (overlap $=.955$)& 40 & 20 & 3 & 3 & 0.70 & 4.4 \\
$\Delta_{\min}=0.036$ (overlap $=.97$) & 40 & 20 & 3 & 3 & 0.70 & 4.5 \\
$\Delta_{\min}=0.015$ (overlap $=1$)   & 40 & 20 & 3 & 3 & 0.70 & 4.5 \\
\multicolumn{7}{l}{\textit{Sparsity}} \\
$\pi=0.05$ & 40 & 20 & 3 & 3 & 0.95 & 4.2 \\
$\pi=0.15$ & 40 & 20 & 3 & 3 & 0.85 & 4.4 \\
$\pi=0.50$ & 40 & 20 & 3 & 3 & 0.50 & 4.5 \\
$\pi=0.80$ & 40 & 20 & 3 & 3 & 0.20 & 4.7 \\
\multicolumn{7}{l}{\textit{Dispersion}} \\
$r=0.2$  & 40 & 20 & 3 & 3 & 0.70 & 16.0 \\
$r=5$    & 40 & 20 & 3 & 3 & 0.70 &  2.4 \\
$r=20$   & 40 & 20 & 3 & 3 & 0.70 &  2.0 \\
$r=200$  & 40 & 20 & 3 & 3 & 0.70 &  1.9 \\
\multicolumn{7}{l}{\textit{Transition persistence}} \\
$\rho=0.40$ & 40 & 20 & 3 & 3 & 0.70 & 4.5 \\
$\rho=0.60$ & 40 & 20 & 3 & 3 & 0.70 & 4.5 \\
$\rho=0.90$ & 40 & 20 & 3 & 3 & 0.70 & 4.5 \\
\multicolumn{7}{l}{\textit{Network dimensions}} \\
$n=80$   &  80 & 20 &  3 & 3 & 0.70 & 4.6 \\
$n=150$  & 150 & 20 &  3 & 3 & 0.70 & 4.6 \\
$T=10$   &  40 & 10 &  3 & 3 & 0.70 & 4.5 \\
$T=45$   &  40 & 45 &  3 & 3 & 0.70 & 4.5 \\
$K=5$    &  40 & 20 &  5 & 3 & 0.70 & 4.5 \\
$K=10$   &  40 & 20 & 10 & 3 & 0.70 & 4.5 \\
\multicolumn{7}{l}{\textit{Regime structure}} \\
Five true regimes ($G=5$) & 40 & 20 & 3 & 5 & 0.70 & 4.5 \\
Rare regime                & 40 & 20 & 3 & 3 & 0.70 & 4.5 \\
Null (no true structure)   & 40 & 20 & 3 & 3 & 0.70 & 8.7 \\
\multicolumn{7}{l}{\textit{Scenario 2: misspecification}} \\
Latent dimension ($d=3\to2$)  & 40 & 20 & 3 & 3 & 0.70 & 3.6 \\
\bottomrule
\end{tabular}
\end{table}


\section{Simulation study: parameter recovery}
\label{sec:S2}
Table~\ref{tab:S2} compares the posterior means of $\alpha_{tk}$, $\beta_{tk}$ and $r_{tk}$ with their true values: bias and RMSE are computed over the $KT$ slices, and coverage is the proportion of slices whose true value lies in the $90\%$ credible interval (reported for the identified contrast $\tilde\alpha_{tk}$, Equation~(2) of the main text); all are averaged over the $20$ replicates.
$\beta_{tk}$, informed by all dyads, has negligible bias and coverage between $0.88$ and $0.91$ everywhere.
The intercept deteriorates with sparsity (bias $0.349$ at $\pi_{tk}=0.05$) and low separation ($-0.233$ at $\Delta_{\min}=0.015$), but coverage of $\tilde\alpha_{tk}$ stays between $0.82$ and $0.90$.
$r_{tk}$ is poorly recovered near the Poisson limit (RMSE $69.7$ at $r_{tk}=20$ and $80.3$ at $r_{tk}=200$), where the likelihood is flat in $r_{tk}$.

\begin{table}[htbp]
\centering
\caption{Recovery of the network-specific parameters. Bias and RMSE of the posterior means over the $KT$ slices; Cov.: proportion of slices whose true value lies in the $90\%$ credible interval (nominal $0.90$), reported for the identified contrast $\tilde\alpha_{tk}$ rather than for $\alpha_{tk}$. Averages over $20$ replicates.}
\label{tab:S2}
\footnotesize
\setlength{\tabcolsep}{2.5pt}
\begin{tabular}{lrrrrrrrrr}
\toprule
 & \multicolumn{3}{c}{$\alpha$} & \multicolumn{3}{c}{$\beta$} & \multicolumn{3}{c}{$r$} \\
\cmidrule(lr){2-4} \cmidrule(lr){5-7} \cmidrule(lr){8-10}
Configuration & Bias & RMSE & Cov.\ ($\tilde\alpha$) & Bias & RMSE & Cov. & Bias & RMSE & Cov. \\
\midrule
\multicolumn{10}{l}{\textit{Regime separation}} \\
$\Delta_{\min}=0.368$ & 0.054 & 0.138 & 0.882 & 0.001 & 0.056 & 0.892 & 0.023 & 0.259 & 0.901 \\
$\Delta_{\min}=0.237$ & 0.052 & 0.145 & 0.859 & 0.001 & 0.055 & 0.901 & 0.009 & 0.278 & 0.883 \\
$\Delta_{\min}=0.166$ & 0.063 & 0.140 & 0.876 & 0.001 & 0.055 & 0.897 & 0.027 & 0.268 & 0.892 \\
$\Delta_{\min}=0.117$ & 0.057 & 0.153 & 0.848 & 0.000 & 0.055 & 0.893 & 0.006 & 0.279 & 0.868 \\
$\Delta_{\min}=0.095$ & 0.047 & 0.148 & 0.848 & 0.003 & 0.055 & 0.890 & -0.001 & 0.271 & 0.881 \\
$\Delta_{\min}=0.057$ & -0.025 & 0.175 & 0.867 & -0.001 & 0.054 & 0.907 & -0.050 & 0.293 & 0.828 \\
$\Delta_{\min}=0.050$ & -0.047 & 0.188 & 0.841 & 0.000 & 0.054 & 0.899 & -0.098 & 0.305 & 0.779 \\
$\Delta_{\min}=0.036$ & -0.113 & 0.233 & 0.829 & -0.003 & 0.055 & 0.902 & -0.134 & 0.316 & 0.736 \\
$\Delta_{\min}=0.015$ & -0.233 & 0.278 & 0.892 & 0.001 & 0.055 & 0.895 & -0.133 & 0.277 & 0.780 \\
\multicolumn{10}{l}{\textit{Sparsity}} \\
$\pi=0.05$ & 0.349 & 0.448 & 0.823 & 0.004 & 0.120 & 0.888 & 1.949 & 6.408 & 0.926 \\
$\pi=0.15$ & 0.103 & 0.220 & 0.848 & 0.000 & 0.069 & 0.899 & 0.039 & 0.411 & 0.911 \\
$\pi=0.50$ & 0.040 & 0.106 & 0.873 & 0.003 & 0.051 & 0.901 & 0.010 & 0.203 & 0.894 \\
$\pi=0.80$ & 0.028 & 0.077 & 0.887 & 0.001 & 0.062 & 0.897 & -0.003 & 0.157 & 0.882 \\
\multicolumn{10}{l}{\textit{Dispersion}} \\
$r=0.2$ & 0.192 & 0.275 & 0.901 & -0.002 & 0.056 & 0.887 & 0.062 & 0.091 & 0.892 \\
$r=5$   & 0.048 & 0.084 & 0.873 & 0.002 & 0.055 & 0.903 & 1.422 & 5.053 & 0.887 \\
$r=20$  & 0.034 & 0.066 & 0.867 & 0.001 & 0.057 & 0.892 & 48.963 & 69.651 & 0.925 \\
$r=200$ & 0.036 & 0.059 & 0.864 & 0.000 & 0.056 & 0.893 & -57.456 & 80.255 & 0.940 \\
\multicolumn{10}{l}{\textit{Transition persistence}} \\
$\rho=0.40$ & 0.060 & 0.136 & 0.892 & -0.002 & 0.055 & 0.904 & 0.011 & 0.262 & 0.901 \\
$\rho=0.60$ & 0.061 & 0.139 & 0.873 & 0.001 & 0.054 & 0.904 & 0.022 & 0.264 & 0.894 \\
$\rho=0.90$ & 0.063 & 0.145 & 0.882 & -0.003 & 0.056 & 0.898 & 0.018 & 0.260 & 0.903 \\
\multicolumn{10}{l}{\textit{Network dimensions}} \\
$n=80$  & 0.035 & 0.067 & 0.892 & 0.000 & 0.027 & 0.907 & -0.007 & 0.122 & 0.906 \\
$n=150$ & 0.021 & 0.038 & 0.897 & 0.000 & 0.014 & 0.903 & -0.001 & 0.066 & 0.897 \\
$T=10$  & 0.125 & 0.201 & 0.855 & 0.001 & 0.055 & 0.892 & 0.007 & 0.273 & 0.893 \\
$T=45$  & 0.013 & 0.120 & 0.889 & -0.001 & 0.056 & 0.896 & 0.012 & 0.259 & 0.895 \\
$K=5$   & 0.032 & 0.122 & 0.901 & 0.002 & 0.055 & 0.901 & 0.018 & 0.260 & 0.901 \\
$K=10$  & -0.006 & 0.123 & 0.889 & -0.001 & 0.055 & 0.904 & 0.018 & 0.264 & 0.892 \\
\multicolumn{10}{l}{\textit{Regime structure}} \\
Five true regimes ($G=5$) & 0.103 & 0.185 & 0.852 & 0.000 & 0.054 & 0.899 & 0.013 & 0.274 & 0.891 \\
Rare regime                & 0.063 & 0.146 & 0.897 & -0.003 & 0.055 & 0.893 & 0.022 & 0.264 & 0.895 \\
Null (no true structure)   & 0.204 & 0.232 & 0.875 & -0.001 & 0.057 & 0.882 & 0.009 & 0.199 & 0.907 \\
\multicolumn{10}{l}{\textit{Scenario 2: misspecification}} \\
Latent dimension ($d=3\to2$) & -0.126 & 0.194 & 0.854 & 0.000 & 0.055 & 0.897 & -0.011 & 0.276 & 0.885 \\
\bottomrule
\end{tabular}
\end{table}


\section{Simulation study: latent-space recovery}
\label{sec:S3}
Table~\ref{tab:S3} extends Figure~1 of the main text to every configuration.
The Procrustes correlation compares aligned estimated and true coordinates, averaged over all $G_{\mathrm{true}}$ regimes; the distance correlation compares pairwise distances and is the minimum over the occupied regimes; coverage of $Z$ is the proportion of true coordinates within the $90\%$ credible intervals.
As separation decreases, the model leaves regimes empty and the Procrustes correlation falls to $0.485$ at $\Delta_{\min}=0.015$, whereas the distance correlation of the occupied regimes stays above $0.97$.
Elsewhere it exceeds $0.92$, except at $\pi_{tk}=0.05$ ($0.890$), and coverage of $Z$ lies between $0.80$ and $0.89$, except when the latent dimension is underspecified ($0.566$).

\begin{table}[htbp]
\centering
\caption{Recovery of the latent configurations. Procrustes corr.: correlation between true and Procrustes-aligned posterior mean coordinates, averaged over all $G_{\mathrm{true}}$ regimes; Dist.\ corr.\ (occ.): correlation between estimated and true pairwise distances, minimum over the occupied regimes; Coverage $Z$: proportion of true coordinates within the $90\%$ credible intervals; Occ.\ regimes: number of regimes with at least one slice in the modal sequence. Averages over $20$ replicates.}
\label{tab:S3}
\setlength{\tabcolsep}{4pt}
\begin{tabular}{lrrrr}
\toprule
Configuration & Procrustes corr. & Dist.\ corr.\ (occ.) & Coverage $Z$ & Occ.\ regimes \\
\midrule
\multicolumn{5}{l}{\textit{Regime separation}} \\
$\Delta_{\min}=0.368$ & 0.995 & 0.988 & 0.865 & 3.00 \\
$\Delta_{\min}=0.237$ & 0.995 & 0.986 & 0.872 & 3.00 \\
$\Delta_{\min}=0.166$ & 0.995 & 0.987 & 0.867 & 3.00 \\
$\Delta_{\min}=0.117$ & 0.980 & 0.986 & 0.853 & 2.95 \\
$\Delta_{\min}=0.095$ & 0.982 & 0.987 & 0.863 & 2.95 \\
$\Delta_{\min}=0.057$ & 0.833 & 0.981 & 0.849 & 2.40 \\
$\Delta_{\min}=0.050$ & 0.802 & 0.975 & 0.835 & 2.30 \\
$\Delta_{\min}=0.036$ & 0.702 & 0.973 & 0.808 & 1.95 \\
$\Delta_{\min}=0.015$ & 0.485 & 0.973 & 0.803 & 1.15 \\
\multicolumn{5}{l}{\textit{Sparsity}} \\
$\pi=0.05$ & 0.954 & 0.890 & 0.826 & 3.00 \\
$\pi=0.15$ & 0.987 & 0.967 & 0.853 & 3.00 \\
$\pi=0.50$ & 0.997 & 0.994 & 0.877 & 3.00 \\
$\pi=0.80$ & 0.997 & 0.993 & 0.876 & 3.00 \\
\multicolumn{5}{l}{\textit{Dispersion}} \\
$r=0.2$ & 0.994 & 0.984 & 0.885 & 3.00 \\
$r=5$   & 0.995 & 0.990 & 0.859 & 3.00 \\
$r=20$  & 0.992 & 0.982 & 0.826 & 3.00 \\
$r=200$ & 0.993 & 0.989 & 0.826 & 3.00 \\
\multicolumn{5}{l}{\textit{Transition persistence}} \\
$\rho=0.40$ & 0.996 & 0.991 & 0.861 & 3.00 \\
$\rho=0.60$ & 0.996 & 0.991 & 0.864 & 3.00 \\
$\rho=0.90$ & 0.993 & 0.979 & 0.860 & 3.00 \\
\multicolumn{5}{l}{\textit{Network dimensions}} \\
$n=80$  & 0.998 & 0.994 & 0.873 & 3.00 \\
$n=150$ & 0.999 & 0.997 & 0.890 & 3.00 \\
$T=10$  & 0.973 & 0.958 & 0.857 & 2.95 \\
$T=45$  & 0.998 & 0.996 & 0.863 & 3.00 \\
$K=5$   & 0.997 & 0.994 & 0.861 & 3.00 \\
$K=10$  & 0.999 & 0.997 & 0.879 & 3.00 \\
\multicolumn{5}{l}{\textit{Regime structure}} \\
Five true regimes ($G=5$) & 0.944 & 0.922 & 0.868 & 4.75 \\
Rare regime                & 0.963 & 0.973 & 0.866 & 2.90 \\
Null (no true structure)   & 0.490 & 0.988 & 0.880 & 1.15 \\
\multicolumn{5}{l}{\textit{Scenario 2: misspecification}} \\
Latent dimension ($d=3\to2$) & 0.978 & 0.980 & 0.566 & 3.00 \\
\bottomrule
\end{tabular}
\end{table}


\section{Simulation study: regime and transition-matrix recovery}
\label{sec:S4}
Table~\ref{tab:S4} extends Figure~3 of the main text.
ARI compares the modal and true regime sequences ($1$: identical partitions; $0$: chance agreement); the transition probabilities, computed draw by draw from $\boldsymbol{\Gamma}$ and $\boldsymbol{\Lambda}$, are compared with the true ones through the mean absolute error, the coverage of the $90\%$ credible intervals and their width, averaged over entries and layers.
ARI is at least $0.99$ everywhere except at $\Delta_{\min}\le0.117$ and in the null configuration, so regime recovery depends essentially on separation.
The transition probabilities have errors of about $0.10$ and depend on the number of observed transitions, improving with $T=45$ (error $0.060$) and worsening with $T=10$ ($0.144$); coverage falls to $0.61$--$0.66$ when the regimes are not identified ($\Delta_{\min}=0.015$ and null configuration).

\begin{table}[htbp]
\centering
\caption{Recovery of the regime sequences and transition matrices. ARI: adjusted Rand index between modal and true regime sequences; MAE, Cov.\ and Width $\widehat P$: mean absolute error, coverage of the $90\%$ credible intervals and their average width for the transition probabilities, averaged over entries and layers; Occ.\ regimes: number of regimes with at least one slice; Min.\ occ.: smallest share of slices assigned to a regime. Averages over $20$ replicates.}
\label{tab:S4}
\setlength{\tabcolsep}{4pt}
\begin{tabular}{lrrrrrr}
\toprule
Configuration & ARI & MAE $\widehat P$ & Cov.\ $\widehat P$ & Width $\widehat P$ & Occ.\ regimes & Min.\ occ. \\
\midrule
\multicolumn{7}{l}{\textit{Regime separation}} \\
$\Delta_{\min}=0.368$ & 1.000 & 0.102 & 0.889 & 0.371 & 3.00 & 0.205 \\
$\Delta_{\min}=0.237$ & 1.000 & 0.086 & 0.928 & 0.362 & 3.00 & 0.212 \\
$\Delta_{\min}=0.166$ & 1.000 & 0.096 & 0.920 & 0.368 & 3.00 & 0.212 \\
$\Delta_{\min}=0.117$ & 0.971 & 0.115 & 0.865 & 0.378 & 2.95 & 0.201 \\
$\Delta_{\min}=0.095$ & 0.978 & 0.114 & 0.857 & 0.378 & 2.95 & 0.196 \\
$\Delta_{\min}=0.057$ & 0.747 & 0.149 & 0.806 & 0.432 & 2.40 & 0.101 \\
$\Delta_{\min}=0.050$ & 0.698 & 0.153 & 0.813 & 0.440 & 2.30 & 0.104 \\
$\Delta_{\min}=0.036$ & 0.490 & 0.191 & 0.739 & 0.488 & 1.95 & 0.054 \\
$\Delta_{\min}=0.015$ & 0.079 & 0.245 & 0.661 & 0.594 & 1.15 & 0.000 \\
\multicolumn{7}{l}{\textit{Sparsity}} \\
$\pi=0.05$ & 0.991 & 0.105 & 0.885 & 0.375 & 3.00 & 0.202 \\
$\pi=0.15$ & 1.000 & 0.108 & 0.881 & 0.369 & 3.00 & 0.174 \\
$\pi=0.50$ & 1.000 & 0.108 & 0.883 & 0.376 & 3.00 & 0.212 \\
$\pi=0.80$ & 1.000 & 0.107 & 0.887 & 0.371 & 3.00 & 0.197 \\
\multicolumn{7}{l}{\textit{Dispersion}} \\
$r=0.2$ & 1.000 & 0.112 & 0.878 & 0.378 & 3.00 & 0.199 \\
$r=5$   & 1.000 & 0.099 & 0.898 & 0.366 & 3.00 & 0.210 \\
$r=20$  & 1.000 & 0.118 & 0.857 & 0.386 & 3.00 & 0.181 \\
$r=200$ & 1.000 & 0.101 & 0.896 & 0.371 & 3.00 & 0.229 \\
\multicolumn{7}{l}{\textit{Transition persistence}} \\
$\rho=0.40$ & 1.000 & 0.104 & 0.902 & 0.401 & 3.00 & 0.259 \\
$\rho=0.60$ & 1.000 & 0.105 & 0.872 & 0.391 & 3.00 & 0.247 \\
$\rho=0.90$ & 1.000 & 0.112 & 0.843 & 0.359 & 3.00 & 0.158 \\
\multicolumn{7}{l}{\textit{Network dimensions}} \\
$n=80$  & 1.000 & 0.098 & 0.917 & 0.372 & 3.00 & 0.194 \\
$n=150$ & 1.000 & 0.109 & 0.885 & 0.370 & 3.00 & 0.193 \\
$T=10$  & 1.000 & 0.144 & 0.870 & 0.482 & 2.95 & 0.150 \\
$T=45$  & 1.000 & 0.060 & 0.924 & 0.252 & 3.00 & 0.225 \\
$K=5$   & 1.000 & 0.085 & 0.909 & 0.333 & 3.00 & 0.238 \\
$K=10$  & 1.000 & 0.079 & 0.885 & 0.297 & 3.00 & 0.258 \\
\multicolumn{7}{l}{\textit{Regime structure}} \\
Five true regimes ($G=5$) & 0.998 & 0.104 & 0.911 & 0.370 & 4.75 & 0.056 \\
Rare regime                & 1.000 & 0.118 & 0.874 & 0.388 & 2.90 & 0.101 \\
Null (no true structure)   & -0.001 & 0.262 & 0.609 & 0.593 & 1.15 & 0.000 \\
\multicolumn{7}{l}{\textit{Scenario 2: misspecification}} \\
Latent dimension ($d=3\to2$) & 1.000 & 0.103 & 0.902 & 0.380 & 3.00 & 0.198 \\
\bottomrule
\end{tabular}
\end{table}


\section{Simulation study: network-level recovery}
\label{sec:S5}
Table~\ref{tab:S5} compares the estimated intensities $\hat\mu_{tk,ij}=\exp\{\hat\alpha_{tk}-\hat D_{\hat S_{tk}}(i,j)\}$ with the true ones over all dyads (MAE and RMSE $\mu$) and with the observed positive counts (MAE$^+$, RMSE$^+$, logMAE$^+$).
Errors against the data are about seven times larger than against the truth at baseline ($1.722$ against $0.251$), because they include the sampling variability of each count and because $\hat\mu_{tk,ij}$ is the mean of the untruncated component, smaller than the expected positive count (Section~2.3.1 of the main text).
Errors against the truth increase as separation decreases ($0.360$ at $\Delta_{\min}=0.015$) and are largest in sparse ($0.871$ at $\pi_{tk}=0.05$) and strongly overdispersed ($0.473$ at $r_{tk}=0.2$) networks.

\begin{table}[htbp]
\centering
\caption{Network-level recovery. MAE and RMSE $\mu$: errors of the estimated intensities $\hat\mu_{tk,ij}$ against the true ones, over all directed dyads; MAE$^+$, RMSE$^+$, logMAE$^+$: errors of $\hat\mu_{tk,ij}$ against the observed positive counts, on the original and log scale. Averages over $20$ replicates.}
\label{tab:S5}
\begin{tabular}{lrrrrr}
\toprule
Configuration & MAE $\mu$ & RMSE $\mu$ & MAE$^+$ & RMSE$^+$ & logMAE$^+$ \\
\midrule
\multicolumn{6}{l}{\textit{Regime separation}} \\
$\Delta_{\min}=0.368$ & 0.251 & 0.513 & 1.722 & 2.930 & 1.300 \\
$\Delta_{\min}=0.237$ & 0.249 & 0.507 & 1.699 & 2.894 & 1.327 \\
$\Delta_{\min}=0.166$ & 0.245 & 0.504 & 1.724 & 2.931 & 1.333 \\
$\Delta_{\min}=0.117$ & 0.251 & 0.515 & 1.706 & 2.921 & 1.341 \\
$\Delta_{\min}=0.095$ & 0.249 & 0.512 & 1.702 & 2.902 & 1.337 \\
$\Delta_{\min}=0.057$ & 0.290 & 0.593 & 1.701 & 2.913 & 1.342 \\
$\Delta_{\min}=0.050$ & 0.320 & 0.638 & 1.705 & 2.950 & 1.320 \\
$\Delta_{\min}=0.036$ & 0.347 & 0.692 & 1.738 & 3.017 & 1.380 \\
$\Delta_{\min}=0.015$ & 0.360 & 0.695 & 1.733 & 2.989 & 1.384 \\
\multicolumn{6}{l}{\textit{Sparsity}} \\
$\pi=0.05$ & 0.871 & 1.773 & 1.778 & 3.007 & 1.185 \\
$\pi=0.15$ & 0.408 & 0.811 & 1.724 & 2.940 & 1.230 \\
$\pi=0.50$ & 0.185 & 0.379 & 1.724 & 2.945 & 1.343 \\
$\pi=0.80$ & 0.143 & 0.292 & 1.763 & 3.018 & 1.330 \\
\multicolumn{6}{l}{\textit{Dispersion}} \\
$r=0.2$ & 0.473 & 0.939 & 3.262 & 7.962 & 1.360 \\
$r=5$   & 0.161 & 0.326 & 1.276 & 1.752 & 1.221 \\
$r=20$  & 0.132 & 0.269 & 1.151 & 1.480 & 1.205 \\
$r=200$ & 0.125 & 0.240 & 1.103 & 1.387 & 1.183 \\
\multicolumn{6}{l}{\textit{Transition persistence}} \\
$\rho=0.40$ & 0.258 & 0.512 & 1.724 & 2.934 & 1.307 \\
$\rho=0.60$ & 0.253 & 0.501 & 1.718 & 2.916 & 1.288 \\
$\rho=0.90$ & 0.251 & 0.520 & 1.743 & 2.966 & 1.331 \\
\multicolumn{6}{l}{\textit{Network dimensions}} \\
$n=80$  & 0.150 & 0.312 & 1.752 & 2.988 & 1.341 \\
$n=150$ & 0.101 & 0.210 & 1.748 & 2.980 & 1.348 \\
$T=10$  & 0.365 & 0.756 & 1.719 & 2.904 & 1.263 \\
$T=45$  & 0.187 & 0.381 & 1.737 & 2.972 & 1.359 \\
$K=5$   & 0.203 & 0.410 & 1.733 & 2.956 & 1.323 \\
$K=10$  & 0.175 & 0.355 & 1.727 & 2.953 & 1.367 \\
\multicolumn{6}{l}{\textit{Regime structure}} \\
Five true regimes ($G=5$) & 0.320 & 0.673 & 1.736 & 2.951 & 1.279 \\
Rare regime                & 0.242 & 0.516 & 1.735 & 2.961 & 1.325 \\
Null (no true structure)   & 0.293 & 0.634 & 2.462 & 4.523 & 1.357 \\
\multicolumn{6}{l}{\textit{Scenario 2: misspecification}} \\
Latent dimension ($d=3\to2$) & 0.267 & 0.536 & 1.583 & 2.615 & 1.318 \\
\bottomrule
\end{tabular}
\end{table}


\section{Simulation study: posterior predictive checks}
\label{sec:S6}
Table~\ref{tab:S6} extends Figure~4 of the main text to every configuration.
For each replicate, $1{,}000$ draws of the first chain are used to simulate replicated collections, on which we compute the proportion of zero edges, the mean and variance of the positive counts and the maximum count; bias is the mean replicated value minus the observed one, and coverage the percentage of replicates in which the observed value lies within the $5$--$95\%$ range of the replicated values.
The proportion of zeros is always reproduced, and the mean of the positive counts is covered in $90$--$100\%$ of the replicates except at $\pi_{tk}=0.05$ ($75\%$).
The variance is overestimated in every configuration except $r_{tk}=0.2$, and its coverage is below $50\%$ in most configurations, down to $0\%$ at $\pi_{tk}\le0.15$ and $5\%$ at $r_{tk}=200$.

\begin{table}[htbp]
\centering
\small
\setlength{\tabcolsep}{4pt}
\caption{Posterior predictive checks. Statistics computed on the whole collection: proportion of zero edges, mean and variance of the positive counts, maximum count. Bias: mean replicated value minus observed value; Coverage: percentage of the $20$ replicates in which the observed value lies in the $5$--$95\%$ range of $1{,}000$ replicated values.}
\label{tab:S6}
\begin{tabular}{l rrrr rrr}
\toprule
& \multicolumn{4}{c}{Bias} & \multicolumn{3}{c}{Coverage (\%)} \\
\cmidrule(lr){2-5} \cmidrule(lr){6-8}
Configuration & zeros & mean$^+$ & var$^+$ & max & zeros & mean & var \\
\midrule
\multicolumn{8}{l}{\textit{Regime separation}} \\
$\Delta_{\min}=0.368$ & -0.00 & 0.02 & 0.87 &  6.45 & 100 & 100 & 35 \\
$\Delta_{\min}=0.237$ & -0.00 & 0.02 & 0.94 &  7.33 & 100 & 100 & 35 \\
$\Delta_{\min}=0.166$ & -0.00 & 0.01 & 0.89 &  8.67 & 100 & 100 & 55 \\
$\Delta_{\min}=0.117$ & -0.00 & 0.02 & 0.88 &  8.13 & 100 & 100 & 55 \\
$\Delta_{\min}=0.095$ & -0.00 & 0.01 & 0.93 &  6.91 & 100 &  95 & 45 \\
$\Delta_{\min}=0.057$ & -0.00 & 0.02 & 1.13 &  9.41 & 100 & 100 & 35 \\
$\Delta_{\min}=0.050$ & -0.00 & 0.02 & 1.21 & 11.92 & 100 &  95 & 35 \\
$\Delta_{\min}=0.036$ & -0.00 & 0.02 & 1.20 & 12.03 & 100 &  90 & 35 \\
$\Delta_{\min}=0.015$ & -0.00 & 0.01 & 0.66 &  7.32 & 100 & 100 & 90 \\
\multicolumn{8}{l}{\textit{Sparsity}} \\
$\pi=0.05$ & -0.00 & 0.11 & 5.90 & 36.48 & 100 &  75 &  0 \\
$\pi=0.15$ & -0.00 & 0.04 & 2.04 & 17.38 & 100 &  95 &  0 \\
$\pi=0.50$ & -0.00 & 0.01 & 0.48 &  4.57 & 100 & 100 & 90 \\
$\pi=0.80$ &  0.00 & 0.01 & 0.36 &  7.61 & 100 & 100 & 85 \\
\multicolumn{8}{l}{\textit{Dispersion}} \\
$r=0.2$ & -0.00 &  0.04 & -0.34 & -7.16 & 100 & 100 & 100 \\
$r=5$   & -0.00 &  0.00 &  0.21 &  3.58 & 100 & 100 &  75 \\
$r=20$  & -0.00 &  0.00 &  0.11 &  2.54 & 100 & 100 &  90 \\
$r=200$ & -0.00 & -0.00 &  0.17 &  1.84 & 100 & 100 &   5 \\
\multicolumn{8}{l}{\textit{Transition persistence}} \\
$\rho=0.40$ & -0.00 & 0.02 & 1.01 &  9.22 & 100 & 100 & 20 \\
$\rho=0.60$ & -0.00 & 0.02 & 1.00 & 11.10 & 100 & 100 & 30 \\
$\rho=0.90$ & -0.00 & 0.02 & 1.01 &  7.57 & 100 & 100 & 35 \\
\multicolumn{8}{l}{\textit{Network dimensions}} \\
$n=80$  & -0.00 & 0.01 & 0.36 &  2.26 & 100 &  90 & 60 \\
$n=150$ & -0.00 & 0.00 & 0.17 &  0.27 & 100 &  95 & 70 \\
$T=10$  & -0.00 & 0.03 & 1.50 & 12.49 & 100 & 100 & 20 \\
$T=45$  & -0.00 & 0.01 & 0.63 &  6.30 & 100 & 100 & 35 \\
$K=5$   & -0.00 & 0.01 & 0.75 & 10.39 & 100 & 100 & 25 \\
$K=10$  & -0.00 & 0.01 & 0.54 & 10.72 & 100 & 100 & 10 \\
\multicolumn{8}{l}{\textit{Regime structure}} \\
Five true regimes ($G=5$) & -0.00 & 0.02 & 1.26 & 14.34 & 100 &  95 & 10 \\
Rare regime                & -0.00 & 0.02 & 0.86 &  5.97 & 100 & 100 & 55 \\
Null (no true structure)   & -0.00 & 0.02 & 1.49 &  5.63 & 100 & 100 & 95 \\
\multicolumn{8}{l}{\textit{Scenario 2: misspecification}} \\
Latent dimension ($d=3\to2$) & -0.00 & 0.02 & 0.69 & 9.26 & 100 & 100 & 25 \\
\bottomrule
\end{tabular}
\end{table}


\section{Simulation study: MCMC diagnostics}
\label{sec:S7}
Table~\ref{tab:S7} reports the $\widehat R$ diagnostics for the model parameters, using the rank-normalised split-$\widehat R$ of \citet{vehtari2021rank}, computed with \texttt{posterior::rhat} from the retained draws of the four chains.
Since the chains may label the regimes differently, the regime labels are aligned across chains by matching each chain to a common reference labelling that maximises the agreement of their modal regime sequences; the last column reports the resulting agreement.
The latent configurations enter through the pairwise distances of the occupied regimes, which are invariant to rotations and reflections, rather than through the coordinates; $300$ randomly chosen pairs are used per regime.
All four chains start from the same initial values (Section~3.1 of the main text), so these diagnostics measure agreement among chains started from a common point.
Across the design, the median $\widehat R$ is $1.00$ for every type of parameter, and $\beta_{tk}$ never exceeds $1.1$.
Large values are concentrated in two situations.
At $\mathrm{overlap}=0.955$ and $0.97$, the chains disagree on the regime sequence (agreement $0.92$ and $0.94$), and all regime-dependent parameters are affected, with $6$--$15\%$ of them above $1.1$.
At $r_{tk}=5$, $20$ and $200$ and in the null configuration, the chains agree on the regimes and on $r_{tk}$, but $5$--$20\%$ of the latent distances exceed $1.1$, because the chains settle in different local configurations of the latent positions.

\begin{table}[htbp]
\centering
\caption{Rank-normalised split-$\widehat R$ of the model parameters, computed from the retained draws of the four chains. For each type of parameter: $99$th percentile of $\widehat R$ (q99) and percentage of parameters with $\widehat R>1.1$ (\%), pooled over the parameters of the $20$ replicates. Distances: latent distances of the occupied regimes ($300$ pairs per regime); $P$: transition probabilities; Agr.: mean agreement between the modal regime sequences across the four chains, after aligning the regime labels.}
\label{tab:S7}
\footnotesize
\setlength{\tabcolsep}{3.5pt}
\begin{tabular}{l rr rr rr rr rr r}
\toprule
& \multicolumn{2}{c}{$\alpha_{tk}$} & \multicolumn{2}{c}{$\beta_{tk}$} & \multicolumn{2}{c}{$r_{tk}$} & \multicolumn{2}{c}{Distances} & \multicolumn{2}{c}{$P$}
& \\
\cmidrule(lr){2-3} \cmidrule(lr){4-5} \cmidrule(lr){6-7} \cmidrule(lr){8-9} \cmidrule(lr){10-11}
Configuration & q99 & \% & q99 & \% & q99 & \% & q99 & \% & q99 & \% & Agr. \\
\midrule
\multicolumn{12}{l}{\textit{Regime separation}} \\
$\Delta_{\min}=0.368$ & 1.038 & 0.0 & 1.003 & 0.0 & 1.009 & 0.0 & 1.048 & 0.7 & 1.004 & 0.0 & 1.000 \\
$\Delta_{\min}=0.237$ & 1.008 & 0.0 & 1.004 & 0.0 & 1.008 & 0.0 & 1.007 & 0.0 & 1.003 & 0.0 & 1.000 \\
$\Delta_{\min}=0.166$ & 1.010 & 0.0 & 1.003 & 0.0 & 1.009 & 0.0 & 1.008 & 0.0 & 1.003 & 0.0 & 1.000 \\
$\Delta_{\min}=0.117$ & 1.009 & 0.0 & 1.003 & 0.0 & 1.009 & 0.0 & 1.007 & 0.0 & 1.003 & 0.0 & 1.000 \\
$\Delta_{\min}=0.095$ & 1.010 & 0.0 & 1.003 & 0.0 & 1.010 & 0.0 & 1.007 & 0.0 & 1.003 & 0.0 & 1.000 \\
$\Delta_{\min}=0.057$ & 1.607 & 3.3 & 1.003 & 0.0 & 1.604 & 2.6 & 1.805 & 3.3 & 1.326 & 1.7 & 0.979 \\
$\Delta_{\min}=0.050$ & 1.677 & 14.6 & 1.003 & 0.0 & 1.660 & 9.9 & 1.824 & 15.3 & 1.796 & 13.7 & 0.921 \\
$\Delta_{\min}=0.036$ & 1.694 & 12.6 & 1.003 & 0.0 & 1.767 & 10.5 & 1.827 & 13.6 & 1.604 & 6.5 & 0.943 \\
$\Delta_{\min}=0.015$ & 1.011 & 0.2 & 1.003 & 0.0 & 1.012 & 0.2 & 1.467 & 4.8 & 1.066 & 0.7 & 0.998 \\
\multicolumn{12}{l}{\textit{Sparsity}} \\
$\pi=0.05$ & 1.389 & 3.4 & 1.003 & 0.0 & 1.573 & 2.0 & 1.580 & 3.1 & 1.334 & 3.9 & 0.984 \\
$\pi=0.15$ & 1.019 & 0.2 & 1.004 & 0.0 & 1.011 & 0.2 & 1.224 & 2.2 & 1.055 & 0.7 & 0.998 \\
$\pi=0.50$ & 1.008 & 0.0 & 1.003 & 0.0 & 1.008 & 0.0 & 1.006 & 0.0 & 1.003 & 0.0 & 1.000 \\
$\pi=0.80$ & 1.092 & 0.8 & 1.003 & 0.0 & 1.006 & 0.0 & 1.049 & 0.8 & 1.003 & 0.0 & 1.000 \\
\multicolumn{12}{l}{\textit{Dispersion}} \\
$r=0.2$ & 1.030 & 0.0 & 1.003 & 0.0 & 1.030 & 0.0 & 1.005 & 0.0 & 1.004 & 0.0 & 1.000 \\
$r=5$ & 1.141 & 2.5 & 1.003 & 0.0 & 1.006 & 0.0 & 1.726 & 5.1 & 1.004 & 0.0 & 1.000 \\
$r=20$ & 1.177 & 1.8 & 1.004 & 0.0 & 1.013 & 0.0 & 1.827 & 13.9 & 1.003 & 0.0 & 1.000 \\
$r=200$ & 1.143 & 3.8 & 1.003 & 0.0 & 1.005 & 0.0 & 1.828 & 20.0 & 1.003 & 0.0 & 1.000 \\
\multicolumn{12}{l}{\textit{Transition persistence}} \\
$\rho=0.40$ & 1.009 & 0.0 & 1.003 & 0.0 & 1.009 & 0.0 & 1.008 & 0.0 & 1.003 & 0.0 & 1.000 \\
$\rho=0.60$ & 1.013 & 0.0 & 1.004 & 0.0 & 1.010 & 0.0 & 1.014 & 0.4 & 1.003 & 0.0 & 1.000 \\
$\rho=0.90$ & 1.022 & 0.2 & 1.003 & 0.0 & 1.009 & 0.1 & 1.232 & 2.2 & 1.053 & 0.2 & 0.999 \\
\multicolumn{12}{l}{\textit{Network dimensions}} \\
$n=80$ & 1.007 & 0.0 & 1.003 & 0.0 & 1.006 & 0.0 & 1.005 & 0.0 & 1.004 & 0.0 & 1.000 \\
$n=150$ & 1.006 & 0.0 & 1.004 & 0.0 & 1.006 & 0.0 & 1.004 & 0.0 & 1.003 & 0.0 & 1.000 \\
$T=10$ & 1.011 & 0.0 & 1.004 & 0.0 & 1.010 & 0.0 & 1.015 & 0.0 & 1.003 & 0.0 & 1.000 \\
$T=45$ & 1.044 & 0.0 & 1.003 & 0.0 & 1.009 & 0.0 & 1.062 & 0.9 & 1.003 & 0.0 & 1.000 \\
$K=5$ & 1.019 & 0.0 & 1.003 & 0.0 & 1.010 & 0.0 & 1.129 & 1.1 & 1.004 & 0.0 & 1.000 \\
$K=10$ & 1.027 & 0.0 & 1.003 & 0.0 & 1.010 & 0.0 & 1.048 & 0.7 & 1.003 & 0.0 & 1.000 \\
\multicolumn{12}{l}{\textit{Regime structure}} \\
Five true regimes ($G=5$) & 1.011 & 0.2 & 1.003 & 0.0 & 1.009 & 0.2 & 1.115 & 1.1 & 1.110 & 1.1 & 0.998 \\
Rare regime & 1.008 & 0.0 & 1.003 & 0.0 & 1.008 & 0.0 & 1.008 & 0.0 & 1.004 & 0.0 & 1.000 \\
Null (no true structure) & 1.007 & 0.5 & 1.003 & 0.0 & 1.006 & 0.3 & 1.649 & 12.9 & 1.275 & 5.2 & 0.995 \\
\multicolumn{12}{l}{\textit{Scenario 2: misspecification}} \\
Latent dimension ($d=3\to2$) & 1.012 & 0.0 & 1.003 & 0.0 & 1.012 & 0.0 & 1.056 & 0.7 & 1.004 & 0.0 & 1.000 \\
\bottomrule
\end{tabular}
\end{table}


\section{Simulation study: scalability}
\label{sec:S8}
Table~\ref{tab:S8} reports the computing time for one dataset, measured as the time needed to run its four chains of $20{,}000$ iterations one after the other, and averaged over replicates.
All computations were run on a computing cluster, on nodes with two AMD EPYC 7452 processors (32 cores each, 2.35 GHz) and 256 GB of RAM, running CentOS Linux~7.
Each dataset was analysed on 4 cores with 6 GB of RAM, using R 4.5.3, with the sampler implemented in C++ via \texttt{Rcpp} and \texttt{RcppArmadillo} and parallelised with OpenMP.
Only the network dimensions are varied; all other settings are at the baseline.
Computing time grows roughly linearly with $T$ and $K$, since the number of slice log-likelihoods to evaluate is proportional to $KT$.
It grows faster with $n$: doubling $n$ from $40$ to $80$ multiplies it by about $4.0$, and moving from $80$ to $150$ by about $3.2$, close to the factors $4$ and $3.5$ implied by the $O(n^2)$ cost of evaluating the pairwise distances (Section~3.2 of the main text).

\begin{table}[htbp]
\centering
\caption{Computing time per dataset (four chains of $20{,}000$ iterations, run sequentially), in seconds, averaged over $20$ replicates. Hardware and software details are given in the text.}
\label{tab:S8}
\begin{tabular}{lrrrr}
\toprule
Configuration & $n$ & $T$ & $K$ & Seconds \\
\midrule
$T=10$  &  40 & 10 &  3 &   660.2 \\
Baseline ($\Delta_{\min}=0.368$) & 40 & 20 & 3 &  1354.2 \\
$K=5$   &  40 & 20 &  5 &  2053.6 \\
$T=45$  &  40 & 45 &  3 &  2801.0 \\
$K=10$  &  40 & 20 & 10 &  4051.6 \\
$n=80$  &  80 & 20 &  3 &  5365.0 \\
$n=150$ & 150 & 20 &  3 & 17318.4 \\
\bottomrule
\end{tabular}
\end{table}


\section{Simulation study: misspecification and robustness}
\label{sec:S9}
Table~\ref{tab:S9a} collects, for the latent-dimension configuration of Scenario~2, the measures of regime, latent-space and mean-structure recovery already reported in Tables~\ref{tab:S3}--\ref{tab:S5}.
The regimes are recovered exactly, the latent distances closely, and the errors on the mean structure are comparable to those at baseline ($0.251$ and $0.513$); the cost of this mild misspecification appears only in the coverage of the latent coordinates (Table~\ref{tab:S3}).

\renewcommand{\thetable}{S9a}
\begin{table}[htbp]
\centering
\caption{Scenario~2, underspecified latent dimension. ARI: adjusted Rand index; Dist.\ corr.\ (occ.): correlation between estimated and true distances, minimum over occupied regimes; MAE and RMSE $\mu$: errors of the estimated intensities against the true ones. Averages over $20$ replicates.}
\label{tab:S9a}
\begin{tabular}{lrrrr}
\toprule
Configuration & ARI & Dist.\ corr.\ (occ.) & MAE $\mu$ & RMSE $\mu$ \\
\midrule
Latent dimension ($d=3\to2$) & 1.000 & 0.980 & 0.267 & 0.536 \\
\bottomrule
\end{tabular}
\end{table}


Table~\ref{tab:S9b} reports the fits with a wrong number of regimes: the $20$ datasets of the baseline, rare-regime and null configurations, all generated with $G_{\mathrm{true}}=3$, are fitted with $G=2$, $3$, $4$ and $5$, so that the comparison across values of $G$ is paired.
For each fit we report the DIC of Section~3.4 of the main text, averaged over chains and replicates, the number of regimes occupied by the modal sequence, $\widehat\Delta_{\min}$, computed here from the posterior mean distance matrices, and the number of replicates in which each value of $G$ attains the lowest DIC.
DIC penalises the under-specified fit in every replicate of the baseline and in $19$ of the $20$ replicates of the rare-regime configuration, but is almost identical for $G=3$, $4$ and $5$: the additional regimes are left empty, so they barely change the deviance, and the lowest DIC is spread across these values.
The occupied regimes, instead, are the same whatever the fitted $G$: three on the baseline datasets, three (or two, in the replicates where the rare regime is never visited) on the rare-regime datasets, and one in most replicates of the null configuration, where every regime but one is empty and $\widehat\Delta_{\min}$ is therefore not reported.

\renewcommand{\thetable}{S9b}
\begin{table}[htbp]
\centering
\caption{Paired fits with $G_{\mathrm{fit}}=2,\ldots,5$ of the $20$ datasets of three configurations. DIC: mean over chains and replicates; Occ.\ regimes: mean number of occupied regimes; $\widehat\Delta_{\min}$: regime-separation diagnostic, from posterior mean distance matrices; Lowest DIC: number of replicates in which $G_{\mathrm{fit}}$ attains the lowest DIC.}
\label{tab:S9b}
\begin{tabular}{lrrrrr}
\toprule
Configuration & $G_{\mathrm{fit}}$ & DIC (mean) & Occ.\ regimes & $\widehat\Delta_{\min}$ & Lowest DIC \\
\midrule
Baseline & 2 & 190686.6 & 2.00 & 0.276 & 0 \\
         & 3 & 184449.5 & 3.00 & 0.373 & 6 \\
         & 4 & 184447.7 & 3.00 & 0.373 & 5 \\
         & 5 & 184447.7 & 3.00 & 0.373 & 9 \\
\midrule
Rare regime & 2 & 187813.8 & 2.00 & 0.331 & 1 \\
            & 3 & 183760.6 & 2.90 & 0.365 & 5 \\
            & 4 & 183759.8 & 2.90 & 0.365 & 5 \\
            & 5 & 183763.4 & 2.90 & 0.364 & 9 \\
\midrule
Null & 2 & 191827.2 & 1.00 & -- & 2 \\
     & 3 & 191830.0 & 1.15 & -- & 8 \\
     & 4 & 191827.7 & 1.00 & -- & 7 \\
     & 5 & 191833.3 & 1.15 & -- & 3 \\
\bottomrule
\end{tabular}
\end{table}


\section{Prior settings}
\label{sec:S10}
Table~\ref{tab:S10} reports the prior hyperparameters used in the simulation study and in the two applications.
The simulation priors are the same in all configurations.
In the applications, the prior means and standard deviations of $\alpha_{tk}$ and the scale $\sigma_a$ of the dispersion prior are set on the scale of the observed counts, which are much larger in GDELT than in the European Parliament data; the remaining hyperparameters are shared across all settings.

\renewcommand{\thetable}{S\arabic{table}}\setcounter{table}{9}
\begin{table}[htbp]
\centering
\caption{Prior hyperparameters (means and standard deviations). Under the sender--receiver (SR) extension, $\alpha_{tk}$ has a Normal prior truncated to $(0,\infty)$.}
\label{tab:S10}
\begin{tabular}{lrrrrrrrrr}
\toprule
 & $\sigma_z$ & $\mu_\alpha$ & $\sigma_\alpha$ & $\mu_\beta$ & $\sigma_\beta$ & $\sigma_a$ & $\sigma_\gamma$ & $\sigma_\lambda$ & $\sigma_u=\sigma_v$ \\
\midrule
Simulation  & 3 & 2  & 2 & 0 & 2 & 1 & 2 & 1 & -- \\
GDELT, base & 3 & 5  & 3 & 0 & 2 & 3 & 2 & 1 & -- \\
GDELT, SR   & 3 & 10 & 6 & 0 & 2 & 3 & 2 & 1 & 1 \\
EP, base    & 3 & 2  & 2 & 0 & 2 & 1 & 2 & 1 & -- \\
EP, SR      & 3 & 4  & 4 & 0 & 2 & 1 & 2 & 1 & 1 \\
\bottomrule
\end{tabular}
\end{table}

\section{Real data applications: data summaries and selection of $G$}
\label{sec:Sapp}
Tables~\ref{tab:gdelt_data} and~\ref{tab:ep_data} summarise the two datasets by layer (Sections~5.1 and~5.2 of the main text), and Tables~\ref{tab:gdelt_selection} and~\ref{tab:ep_selection} report the criteria used to select $G$ (Section~3.4 of the main text).

\begin{table}[t]
\centering
\caption{GDELT dataset summary by layer ($n=45$, $T=138$).}
\label{tab:gdelt_data}
\begin{tabular}{lrrr}
\toprule
Layer & Density & Mean $y>0$ & Max \\
\midrule
Verbal cooperation   & 0.904 & 147.0 & 31,345 \\
Material cooperation & 0.655 &  31.3 &  8,439 \\
Verbal conflict      & 0.609 &  42.5 & 13,764 \\
Material conflict    & 0.642 &  48.4 & 20,715 \\
\midrule
Total                & 0.702 &  74.8 & 31,345 \\
\bottomrule
\end{tabular}
\end{table}

\begin{table}[t]
\centering
\small
\caption{GDELT: selection of $G$ for the base and sender--receiver (SR) models. DIC recomputed on the aligned draws and averaged over chains; minimum regime occupancy of the modal sequence; mean pairwise agreement of the modal regime sequences across chains after relabelling; $\widehat R$ of the log-likelihood. Selection rule: among the values of $G$ with $p_D \geq 0$ in every chain, minimum occupancy $> 0.10$ and mean
agreement $> 0.70$, the smallest mean DIC (bold: $G = 3$ for both models).}
\label{tab:gdelt_selection}
\setlength{\tabcolsep}{3pt}
\begin{tabular}{rrrrrrr}
\toprule
& \multicolumn{3}{c}{Base} & \multicolumn{3}{c}{SR} \\
\cmidrule(lr){2-4} \cmidrule(lr){5-7}
$G$ & DIC & Min.\ occ. & Agreement & DIC & Min.\ occ. & Agreement \\
\midrule
2 & 7,745,976 & 0.33 & 0.95 & 7,414,486 & 0.41 & 1.00 \\
\textbf{3} & \textbf{7,736,439} & \textbf{0.23} & \textbf{0.82} & \textbf{7,403,005} & \textbf{0.23} & \textbf{0.87} \\
4 & 7,732,876 & 0.00 & 0.93 & 7,398,267 & 0.05 & 0.73 \\
5 & 7,727,814 & 0.00 & 0.71 & 7,389,682 & 0.00 & 0.86 \\
6 & 7,730,189 & 0.00 & 0.76 & 7,381,774 & 0.00 & 0.89 \\
7 & 7,726,109 & 0.00 & 0.59 & 7,376,512 & 0.02 & 0.67 \\
8 & 7,724,429 & 0.00 & 0.68 & 7,378,962 & 0.00 & 0.71 \\
\bottomrule
\end{tabular}
\end{table}

\begin{table}[t]
\centering
\caption{European Parliament dataset summary by layer ($n=27$, $T=33$ months, July 2019 to June 2022).}
\label{tab:ep_data}
\begin{tabular}{lrrr}
\toprule
Layer & Density & Mean $y>0$ & Max \\
\midrule
Economy         & 0.737 & 11.5 & 136 \\
Environment     & 0.572 & 16.1 & 459 \\
Foreign affairs & 0.724 & 6.4 & 69 \\
Social          & 0.736 & 11.8 & 134 \\
\midrule
Total           & 0.693 & 11.2 & 459 \\
\bottomrule
\end{tabular}
\end{table}

\begin{table}[t]
\centering
\caption{European Parliament: selection of $G$ for the base and sender--receiver (SR) models. DIC; minimum effective number of parameters across chains; minimum regime occupancy of the modal sequence; mean pairwise agreement (Agr.) of the modal sequences after relabelling.
Selection rule: among the values of $G$ with $p_D \geq 0$ in every chain, minimum occupancy $> 0.10$ and mean agreement $> 0.70$, the smallest mean DIC (bold: $G = 2$ base, $G = 3$ SR).}
\label{tab:ep_selection}
\small
\setlength{\tabcolsep}{4pt}
\begin{tabular}{r rrrr rrrr}
\toprule
& \multicolumn{4}{c}{Base} & \multicolumn{4}{c}{SR} \\
\cmidrule(lr){2-5} \cmidrule(lr){6-9}
$G$ & DIC & Min $p_D$ & Occ. & Agr. & DIC & Min $p_D$ & Occ. & Agr. \\
\midrule
2 & \textbf{406{,}759} & \textbf{568} & \textbf{0.153} & \textbf{0.89} & 394{,}429 & 738 & 0.211 & 0.90 \\
3 & 406{,}583 & 752 & 0.116 & 0.65 & \textbf{392{,}667} & \textbf{906} & \textbf{0.188} & \textbf{0.87} \\
4 & 406{,}564 & 931 & 0.071 & 0.54 & 392{,}117 & 1{,}023 & 0.137 & 0.68 \\
5 & 406{,}468 & 1{,}085 & 0.044 & 0.41 & 391{,}299 & 1{,}099 & 0.104 & 0.48 \\
6 & 406{,}433 & 1{,}183 & 0.038 & 0.53 & 391{,}215 & 1{,}237 & 0.072 & 0.64 \\
7 & 406{,}356 & 1{,}245 & 0.019 & 0.52 & 390{,}359 & 1{,}320 & 0.036 & 0.54 \\
8 & 406{,}270 & 1{,}350 & 0.015 & 0.51 & 390{,}510 & 1{,}402 & 0.038 & 0.52 \\
\bottomrule
\end{tabular}
\end{table}

\end{document}